\documentclass[aps, prd, groupedaddress, eqsecnum, nofootinbib, showpacs, preprintnumbers]{revtex4}
\usepackage{graphicx, epsf, amssymb, amsbsy, amsfonts, amssymb, amsmath}
\usepackage[usenames, dvipsnames]{color}

\usepackage{calc}
\usepackage{url}

\newif\ifdraft
\drafttrue
\newif\ifpreprint
\preprinttrue

\newif\iffigs\figstrue

\DeclareFontFamily{U}{rsf}{}
\DeclareFontShape{U}{rsf}{m}{n}{
  <5> <6> rsfs5 <7> <8> <9> rsfs7 <10-> rsfs10}{}
\DeclareMathAlphabet\Scr{U}{rsf}{m}{n}

\def\sect#1{section~{\ref{#1}}}

\def\app#1{appendix~{\ref{#1}}}
\def\eqn#1{eq.~(\ref{#1})}

\def\eqns#1#2{eqs.~(\ref{#1}) and~(\ref{#2})}

\newcommand{\bea}{\begin{eqnarray}}
\newcommand{\eea}{\end{eqnarray}}
\newcommand{\be}{\begin{equation}}
\newcommand{\ee}{\end{equation}}
\newcommand{\non}{\nonumber}

\def\R{{\mathbb R}}

\def\cD{{\Scr D}}

\def\cS{{\Scr S}}

\def\a{\alpha}

\def\d{\delta}

\def\l{\lambda}

\def\o{\omega}

\def\q{\theta}

\def\L{\Lambda}

\newcommand{\Bv}{\boldsymbol{B}}
\newcommand{\Hv}{\boldsymbol{H}}
\newcommand{\Dv}{\boldsymbol{D}}
\newcommand{\Ev}{\boldsymbol{E}}

\def\nn{\nonumber}
\newcommand{\ad}{{\dot{\alpha}}}                           
  
\newcommand {\cO}{{\cal O}}                          

\newcommand{\pa}{\partial}                           
\newcommand{\hf}{\frac12}

\newcommand{\vf}{\varphi}

\newcommand {\cW}{{\cal W}}

\newcommand {\cZ}{{\cal Z}}

\begin{document}

\preprint{SU-ITP-11/41}

\title{Covariant procedures for perturbative non-linear deformation \\
of duality-invariant theories}

\

\

\author{\bf John Joseph M. Carrasco${}^a$, Renata Kallosh${}^a$, and Radu Roiban${}^b$}

\affiliation{\vskip .43cm
${}^a$Stanford Institute for Theoretical Physics and Department of Physics, Stanford University, 
Stanford, CA 94305-4060, USA\\
${}^b$Department of Physics, Pennsylvania State University, University Park, PA 16802, USA
}

\begin{abstract}
We analyze a recent  conjecture regarding the perturbative construction of non-linear deformations 
of all classically duality invariant theories, including ${\cal N}=8$ supergravity.  Starting with an initial quartic deformation, we engineer a procedure that generates a particular non-linear deformation (Born-Infeld) 
of the Maxwell theory.  This procedure requires the introduction of an infinite number of 
modifications to a constraint which eliminates degrees of freedom consistent with the duality and field content of the system. 
We discuss the extension of this procedure to ${\cal N}=1$ and ${\cal N}=2$ supersymmetric theories, 
and comment on its potential to either construct new  supergravity theories with non-linear Born-Infeld type duality, or to constrain 
the finiteness of ${\cal N}=8$ supergravity. 
\end{abstract}

\pacs{04.65.+e, 11.15.Bt, 11.30.Pb, 03.50.-z, 03.50.De \hspace{1cm}}

\maketitle

\pagestyle{plain}
\setcounter{page}{1}

\section{Introduction}

From our first and most familiar gauge-theory, classical electromagnetism, to the theoretical triumph of maximally supersymmetric supergravity in four-dimensions, ${\cal N}=8$ supergravity~\cite{Cremmer:1978km}, we have at our disposal examples of theories whose equations of motion respect a particularly constraining duality invariance: the rotation of the electric field (or its analog) into the magnetic field.   Their covariant actions, however, must transform non-trivially for the classical duality symmetry of the equations of motion to be preserved~\cite{Gaillard:1981rj,Kuzenko:2000uh,Aschieri:2008ns}. Introducing deformations of the action must be undertaken with a certain amount of care if one wishes to maintain this invariance.  If one is able to consistently include such deformations, exciting generalizations of known theories are possible.  Additionally, one would have the ability to introduce counterterms that might otherwise seem to conflict with the known symmetries of duality-invariant theories.   
In this note we will discuss procedures which, starting from a classical action and quantum generated counterterms, allow us to construct a covariant effective action whose equations of motion are invariant under the same duality transformations as the classical action.

Linear duality has been an integral part of supergravity theories since their beginning~\cite{Ferrara:1976iq, Cremmer:1977tt}.
Non-linear duality models, where the action depends on quartic and higher order powers of vector fields, are well known for gauge theories: these models are generalized Born-Infeld theories discovered in refs.~\cite{BI,Schrodinger}, with a supersymmetric version  later constructed in \cite{Cecotti:1986gb}.  Studied extensively in ~\cite{Gaillard:1981rj,Kuzenko:2000uh,Aschieri:2008ns,Gibbons:1995cv,Gaillard:1997rt,Perry:1996mk,Tseytlin:1999dj,Ketov:2001dq,Gates:2001ff}, 
 they have natural supersymmetric generalizations, as reviewed in refs.~\cite{Kuzenko:2000uh,Aschieri:2008ns}.  
Some attempts to construct the supergravity analog of the Born-Infeld models of non-linear duality have been made in ${\cal N}=1$ supergravity, see e.g. refs.~\cite{Ketov:2001dq,Gates:2001ff} but, as of yet, no models with non-linear duality  have ever been constructed for $N\geq 2$ supergravity.  The possibility that there may exist systematic procedures which can generate them is indeed intriguing.

At present, the ultraviolet properties of
${\cal N}=8$  supergravity in $D=4$ are believed to be related, at least in part, to its duality property, i.e. the symmetry of its equations of motion and Bianchi identities under $E_{7(7)}$ transformations.  The UV properties of ${\cal N}=8$  supergravity in $D=4$ have long been studied, starting with the construction of  candidate $L$-loop order counterterms  for $L\geq 3$ \cite{Kallosh:1980fi, Howe:1980th,Howe:1981xy}.  The three-loop UV divergence supported by the $R^4+(\partial F)^4 + R^2(\partial F)^2+ \cdots$ candidate counterterm~\cite{Kallosh:1980fi,Howe:1981xy} was shown by explicit computations~\cite{Bern:2007hh} to be absent. One set of explanations for this is based on $E_{7(7)}$ symmetry~\cite{Brodel:2009hu, Bossard:2010bd,Beisert:2010jx}. 
$E_{7(7)}$-invariant non-BPS candidate on-shell counterterms with non-linear supersymmetry appear starting at the 8-loop order~\cite{Howe:1980th,Kallosh:1980fi}  and a 1/8 BPS $E_{7(7)}$ candidate counterterm is available at the 7-loop order~\cite{ Bossard:2011tq}.
%
%

From a different perspective, it has been argued~\cite{Kallosh:2010kk} that locality forbids all counterterms in the real light-cone superspace; this provides an alternative explanation of the result of the three-loop computation and an argument in favor of all loop finiteness of ${\cal N}=8$ supergravity. Through a pure spinor worldline formalism, manifest maximal supersymmetry gives another explanation of the three-loop UV finiteness, but suggests a 7-loop four-dimensional divergence~\cite{spinorString}, similar to its string theory counterpart~\cite{Green:2010sp}. 
%

Recently an argument for the all-loop order UV finiteness  of perturbative ${\cal N}=8$ supergravity, in explanation of observed cancellations~\cite{BDR,Bern:2007hh,FourLoop}, was presented in ref.~\cite{Kallosh:2011dp}  based on  the conservation of the Noether-Gaillard-Zumino (NGZ) $E_{7(7)}$ duality current~\cite{Gaillard:1981rj}.  
As we will review in later sections, conservation of the duality current requires the action to transform
 in a specific way. 
 The argument of ref.~\cite{Kallosh:2011dp} is based on the observation that a 
 deformation of the classical ${\cal N}=8$ supergravity action by an $E_{7(7)}$-invariant 
 counterterm leads to an action with different transformation properties and thus to 
 a violation of the $E_{7(7)}$ NGZ current conservation.

It was suggested, however, by Bossard and Nicolai~\cite{BN}, based on previous work on dualities~\cite{Bossard:2010dq, Bunster:2011aw}, 
that  there exist procedures which always allow a duality-consistent perturbative non-linear deformation of general theories --  including ${\cal N}=8$ supergravity -- which exhibit duality-invariant classical equations of motion. An elegant covariant procedure  is described that allows a  nonlinear deformation of classical electromagnetism through a modification of the linear vector field self-duality constraint.  This constraint exists to eliminate degrees of freedom to comply with the field content of the theory and to avoid a double counting of vector fields.   We find that this procedure, at least unmodified, does not reproduce another simple nonlinear deformation of classical electromagnetism: the Born-Infeld theory~\cite{BI,Schrodinger}.   By actively expanding the known Born-Infeld deformation, we are able to {\it a posteriori} derive a procedure that does reproduce it.   We formulate a procedure general enough to find such deformations.  For $U(1)$ theories the deformation is external -- i.e. it may be generated by interactions outside Maxwell's theory.  In interacting theories it is generated by the interactions of the fields of the theory and may either be the result of finite or divergent counterterms.  The procedure we propose has the potential to exclude counterterms that are incompatible with various expectations of the form of the final action.

Extensive analysis suggests that manifestly duality-invariant local actions are not available in the presence of 
Lorentz invariance\footnote{However, Pasti-Soroki-Tonin actions \cite{Pasti:1996vs} are available, which are Lorentz covariant and duality invariant due to a special choice of gauge 
symmetries and a non-polynomial (e.g. inverse powers) dependence on auxiliary fields. In particular, there is an action of this kind  with manifest duality for maximally supersymmetric D=6  supergravity \cite{DePol:2000re}.}.
Manifestly duality-invariant actions with hidden Lorentz invariance were initially constructed for two-dimensional scalar fields in~\cite{Tseytlin:1990nb,Tseytlin:1990va} based on ideas described in~\cite{Floreanini:1987as}\footnote{The ideas of ~\cite{Floreanini:1987as} have also been used in \cite{Henneaux:1988gg} for the construction of actions for self-dual form fields in $2\,{\rm mod}\,4$ dimensions. }.  The generalization of duality-symmetric actions for vector fields in four dimensions (as well as $m$-forms in $d$-dimentions) was explicitly discussed in \cite{Schwarz:1993vs}. 
While Lorentz invariance of the manifestly duality-invariant actions is hidden, it emerges on shell at the classical level and, assuming absence of anomalies, will also be visible at the level of the quantum scattering matrix.  Thus, in such a formulation, the scattering matrix may be expected to be constrained by both manifest  Lorentz and duality invariance.\footnote{In the context of the ${\cal N}=8$ supergravity, certain aspects of the $E_{7(7)}$ duality may be probed at the level of the scattering matrix through soft scalar limits \cite{ArkaniHamed:2008gz}.} 
Analyzing the duality invariance of the effective equations of motion of a covariant formulation of these theories, as we will do in this paper, may be interpreted as an intermediate step towards an analysis of the scattering matrix.

Ref.~\cite{BN} also proposes an explicit  non-covariant construction of duality-invariant theories using the Henneaux-Teitelboim formulation \cite{Henneaux:1988gg,Bunster:2011aw}. In our paper, for the examples limited to the non-linear deformations of the Maxwell theory, we will also discuss the Hamiltonian approach to the problem which has a simple relation to the covariant solution.

We should spend a few words on terminology.  Maxwell theories have no interaction, so the introduction of a non-linear deformation is, of course, a choice.  In supergravity theories, on the other hand, ``experimentally'' identified counterterms (i.e. counterterms arising from explicit calculations) may force deformations upon us.   We will use the word {\it counterterm} to specifically mean  changes to the action necessitated by explicit calculation (or conjectured explicit calculation).  In general the form of a given counterterm will not alone be sufficient to deform the action in a way consistent with the duality.  The procedures discussed in this paper  will generate from these counterterms a final deformed action compatible with duality symmetries.
In Maxwell theories, the role of supergravity counterterms is taken by {\it initial deformation sources} generated by external interactions. Analogously to supergravity theories, the procedures discussed in later sections will take these initial sources and generate final deformed actions.   

The paper is organized as follows. In \sect{dualityEM} we introduce the simplest examples of duality invariant theories, Maxwell's electromagnetism and two of its non-linear deformations.  In \sect{twistedSelfDualConstraintSection} , we introduce  constraints designed to help make duality symmetry manifest, and which allow a framework for introducing deformation.   In \sect{BNprop} we introduce the necessary generalization to supergravity, and reproduce the procedure of ref.~\cite{BN}, for generating non-linear deformations but in notation we will find it easier to generalize from.   In \sect{genProcSection} we derive the procedure required to introduce the Born-Infeld deformation.  In \sect{SUSY} we discuss the applicability of these procedures in a supersymmetric context.  We conclude in \sect{Discussion}. In appendix A we discuss duality in supergravity and in appendix B we present the Hamiltonian solutions of the duality invariant BN and BI models.

\section{Maxwell Duality-Invariant Theories}
\label{dualityEM}
For an excellent review of duality rotations in non-linear electrodynamics, which in this section we follow closely, please see ref.~\cite{Aschieri:2008ns}.  We begin by considering perhaps the most familiar duality-invariant theory, classical electromagnetism in a vacuum.  Maxwell's equations are given
\bea
\label{maxwellEqns}
\partial_t \Bv&=&-\nabla\times \Ev \, ~~,~~~  \,  \, \nabla \cdot\Bv=0\label{max2}
\\
\partial_t \Dv&=&\nabla\times \Hv \,\; ~~~,~~~  \,  \, \nabla \cdot\Dv=0\label{max1} \, \nn
\eea
in addition to  relations between the electric field $\Ev$, the magnetic field $\Hv$, the electric displacement $\Dv$, and the magnetic induction $\Bv$.  In a vacuum, $\Dv=\Ev$, and $\Hv=\Bv$.   The Hamiltonian ${\cal H}=\frac{1}{2}({\bf E}^2+{\bf B}^2)$ and the equations of motion are invariant under rotations 
\be
\left(
                      \begin{array}{cc}
                  \Ev \\
                    \Bv \\
                      \end{array}
                    \right) \mapsto 
 \left(
 \begin{array}{cc}
                  \cos\alpha &  - \sin\alpha\\
                  \sin\alpha & \cos\alpha
                      \end{array}
                    \right) 
                    \left(
                      \begin{array}{cc}
                  \Ev \\
                    \Bv \\
                      \end{array}
 \right)  \, . 
\ee  
 Note that the Lagrangian, however, ${\cal L}={1\over 2}(\Ev^2-\Bv^2)$ is not invariant, for  small rotations $\alpha$ one finds that it transforms as 
\be\delta {\cal L}= -\alpha\,  \Ev \Bv\,.
\ee  
This suggests that non-linear deformations of ${\cal L}$ will require modifications which are also non-invariant.   Indeed the most straightforward non-linear modification is the introduction of a chargeless medium.  In such a medium we will now have non-linear relations:
\be
\label{maxwellMediumRelations}
\Dv=\Dv(\Ev, \Bv)\, \hskip.5cm \Hv=\Hv(\Ev, \Bv)\, .
\ee

It is  convenient to continue the discussion more covariantly through the introduction of four-component notation.  Quite generally, duality transformations may be realized in the path integral as a Legendre transform (see also, e.g.~\cite{Gaillard:1997rt}).  Given some Lagrangian ${\cal L}(F)$ depending only on the field strength of a vector field, one 
constructs
\be
{\widetilde {\cal L}}(F, G) = {\cal L}(F)-\frac{1}{2}\epsilon^{\mu\nu\rho\sigma}F_{\mu\nu}
\partial_\rho{\tilde A}_\sigma~, 
\label{fulldual}
\ee
in which $F$ is treated as a fundamental field. On the one hand, integrating out ${\tilde A}_\sigma$ one finds that $F$ should obey the Bianchi identity $\epsilon^{\mu\nu\rho\sigma}\partial_\nu F_{\rho\sigma}=0$, i.e. that $F$ may be expressed in terms of a vector potential in the usual way. Plugging this into ${\widetilde {\cal L}}(F, G)$ one finds that it reduces to the original Lagrangian ${\cal L}(F)$. 
On the other hand, the classical equations of motion for $F$ require that $G_{\mu\nu}=\partial_\mu{\tilde A}_\nu-\partial_\nu{\tilde A}_\mu$  is related to $F$ by
\be
\label{covConstraint}
\tilde{G}^{\mu \nu} = 2 \frac{ \partial {\cal L}(F)}{\partial F_{\mu \nu}}\,,
\ee
through
\be
\label{gDefs}
G_{\mu\nu}=-{1\over 2}\epsilon_{\mu\nu\rho\sigma}{\tilde G^{\rho\sigma}}
\quad,\qquad
 {{\tilde{G}}}^{\mu\nu}= \frac{1}{2}\epsilon^{\mu\nu\rho\sigma} G_{\rho\sigma} \ .
\ee
The Lagrangian ${\cal L}^D(G)$, dual to ${\cal L}(F)$, is obtained by  eliminating $F$ between 
equations (\ref{covConstraint}) and (\ref{fulldual}). 
Regardless of the form of the original Lagrangian, the Bianchi identity and the equations 
of motion of the original Lagrangian, expressed in terms of $F$ and $G$, are 
\be
\label{covMaxwell}
{\partial}_{\mu}
{\tilde F}^{\mu\nu}  =0~~, \, \hskip.5cm
{\partial}_{\mu}
\tilde{G}^{\mu\nu}=0\, ,
\ee 
and are formally mapped into linear combinations of themselves by a $GL(2)$ transformation. 
Further requiring that the transformed $G$ may be obtained from  the action evaluated on the 
transformed $F$ though eq.~(\ref{covConstraint}) and that the resulting action is a deformation 
of Maxwell's theory ${\cal L}=-\frac{1}{4}F^2+{\cal O}(F^4)$ restricts \cite{Aschieri:2008ns} 
the possible transformations to
\be
\label{constraintMaxwell}
\delta \left(
                      \begin{array}{cc}
                  F \\
                  G \\
                      \end{array}
                    \right)\ =\left(
                      \begin{array}{cc}
                     0 &  B \\
                   -B  &0 \\
                      \end{array}
                    \right)  \left(
                      \begin{array}{cc}
                  F \\
                    G \\
                      \end{array}
                    \right) \, . 
\ee

In other words, the duality transformation exchanges the Bianchi identity and the equations of motion
of the original Lagrangian.
The original Lagrangian is self-dual if ${\cal L}$ and ${\cal L}^D$ have the same functional form. It is easy 
to check that Maxwell's theory, with ${\cal L}(F)=-\frac{1}{4}F^2$, is such a theory. 

In the derivation above, the dual field strength is determined by eq.~(\ref{covConstraint}) and is not an independent field. Since duality transformations (\ref{constraintMaxwell}) mix the field strength and its
dual, it is convenient to interpret $G$ as an independent field and relate it to $F$ by introducing constraint equations as we discuss in \sect{twistedSelfDualConstraintSection}.

For theories with $n_v$ vector fields the strategy for constructing the dual Lagrangian is unchanged. 
The equations of motion and the Bianchi identities remain of the form (\ref{covMaxwell}) but are invariant under a much larger set of transformations:
\be
\label{symplectic}
\delta \left(
                      \begin{array}{cc}
                  F \\
                  G \\
                      \end{array}
                    \right)\ =\left(
                      \begin{array}{cc}
                      A&  B \\
                        C & D \\
                      \end{array}
                    \right)  \left(
                      \begin{array}{cc}
                  F \\
                    G \\
                      \end{array}
                    \right) \, , 
\ee
\be
A^{\mathsf{T}} =-D   \qquad B^{\mathsf{T}}= B \, \qquad C^{\mathsf{T}}=C \, 
\label{T}\ee
 Here $A,  B, C, D$ are the infinitesimal parameters of the transformations, arbitrary real $n\times n$ matrices 
 and the transformations (\ref{symplectic}) generate the $Sp(2n_v, \R)$ algebra. For more general theories, when scalar fields are present, 
 we would also include a   $\delta \phi(A,B,C,D)$.

Consistency of the duality constraint can be expressed as requiring that the Lagrangian must transform under duality in a particular way, defined  by the Noether-Gaillard-Zumino (NGZ) identity~\cite{Gaillard:1981rj}. 
The NGZ current conservation requires universally\footnote{Here we discuss theories with actions depending on  the field strength $F$ but not on its derivatives.
When derivatives are present, an analogous relation is given by a functional derivative over $F$ of the action, see Appendix A.} that for any duality group embeddable into $Sp(2n_v, \R)$
\be
\delta{\cal L}= {1\over 4}  (\tilde G B G+ \tilde FC F )\,.
\label{deltaL}\ee
This leads to  the NGZ identity since the variation $\delta{\cal L}(F, \phi)$ can be computed independently using the chain rule and the information about  $\delta F$ and $\delta \phi$.  

For example, in the case of a  U(1) duality (\ref{constraintMaxwell}),
\be
A=D=0, \hskip 1cm C=-B\,,
\ee 
we see that \eqn{deltaL} reduces to $\delta{\cal L}= {1\over 4}  (\tilde G B G- \tilde FB F ) $. Taking into account that in the absence of scalars 
\be
\delta{\cal L}(F)= \frac{ \partial {\cal L}(F)}{\partial F_{\mu \nu}} \delta F_{\mu \nu}= {1\over 2} \tilde G B G\,,
\ee 
the NGZ identity which follows from 
(\ref{deltaL}) requires that
\be
{1\over 2} \tilde G B G= {1\over 4}  (\tilde G B G- \tilde FB F )\,.
\label{deltaL1}\ee
In this case the NGZ identity simplifies to the following  relation
 \be
 \label{consistencyEqn}
 F\tilde{F} + G\tilde{G} =0\, .
 \ee
 
 \newpage
  The NGZ identity can be alternatively be presented by as follows.  First consider the generalization of the action~(\ref{fulldual}) to the presence of scalars, ${\cal L}(F)\mapsto {\cal L}(F, \phi)$ written in terms of the dual field strength ${\widetilde{\cal L}}(F, \phi)={\cal L}(F, \phi)-\frac{1}{4}F{\tilde G}$.  Now we consider its invariance under duality transformations  (\ref{symplectic}) and $\delta \phi$.  Annotating the transformed $F,\tilde G$ as $F', \tilde G'$, and the transformed $\phi$ as a $\phi'$, 
the invariance of this action implies that   
\be
\int {\widetilde{\cal L}}(F, \phi)=S_{\rm inv} 
=  S[F', \phi'] -  {1\over 4} \int  F' \tilde G' =S[F, \phi] -  {1\over 4} \int  F \tilde G \ .
\label{sinv}
\ee
According to (\ref{symplectic}), (\ref{T})
\be
\delta ( F\tilde G )= (AF  + BG)\tilde G + F(C\tilde F + D\tilde G)=\tilde G B G+ \tilde FC F \ ,
\label{linearvar}
\ee
implying that $S_{\rm inv}$ is invariant under the transformations  (\ref{symplectic}), 
provided that (\ref{deltaL}) is satisfied.

We may also present the NGZ identity as follows  
 \be
 \tilde G -     F \, {\delta  \tilde G\over \delta F} = 4 \, {\delta S_{\rm inv}\over \delta F} \ ,
\label{reconstr}
\ee
which is just the derivative of the defining relation of $S_{\rm inv}$ with respect to $F$ under the assumption that there is some relation between $F$ and $G$.
We can call it a ``reconstruction identity'' since it follows from the form of the action 
\be 
S={1\over 4} \int  F {\tilde G} +S_{\rm inv}\,
\label{actionForm}
\ee  
reconstructed using the duality symmetry.  When the theory only has linear duality (e.g. only $F^2$ terms in the action) $\delta S_{\rm inv} / \delta F$ vanishes.   So \eqns{actionForm}{reconstr}  tell us that any higher order dependence ($F^4$, $F^6$ etc.) must be part of $S_{\rm inv}$.

The NGZ identity, in conjunction with \eqn{covConstraint} can be  solved to find $G(F)$ and various Lagrangians providing a duality symmetry between equations of motion and Bianchi identities. We will discuss two cases of  non-linear deformations of the Maxwell theory for models depending only on $F$'s without derivatives.

\subsection{Born-Infeld Lagrangian}
\label{BILagrangianSection}

The Born-Infeld Lagrangian, perhaps the most venerable non-linear deformation of Maxwell's theory, is
\be
 \label{BILag}
 {\cal L}_{\rm BI}=g^{-2}(1 - \sqrt{\Delta})= - \frac{1}{4} F^2 
 + \frac{1}{32} g^2 \left((F^2)^2 +(F{\tilde F})^2\right)+\cdots \,,
\ee
where $g$ is the coupling constant,  and $\Delta = 1 + 2 g^2 (F^2/4) - g^4 (F \tilde{F}/4)^2$.   Using \eqn{covConstraint}, we
find the following expression for $G$ ,
\bea
\label{GsolveBI}
G_{\mu \nu}&=&-\epsilon_{\mu\nu\rho\sigma}  \frac{ \partial {\cal L}(F)}{\partial F_{\rho \sigma}}\\
&=&\frac{1
 }{\sqrt{\Delta}}(\tilde{F}_{\mu \nu} + g^2 \frac{1}{4}( F \tilde{F}) F_{\mu \nu})\,.
\eea
A little algebra shows that the NGZ identity \eqn{consistencyEqn} is readily verified and that the dual Lagrangian constructed as described above has the same functional form as ${\cal L}_{\rm BI}$.   It is worth noting that classical electromagnetism corresponds to $g^2\to0$.

For relative compactness, and to compare this Lagrangian with the next deformed theory, we introduce the following notation for the two possible Lorentz invariants,
\be
\label{tzDef}
 t=\frac{1}{4} F^2 \quad , \qquad z=\frac{1}{4} F \tilde{F} \ .
\ee
With these field variables, one can rewrite the Born-Infeld Lagrangian simply as
\be
{\cal L}_{\rm BI}=g^{-2}(1 - \sqrt{1 + 2 g^2 t - g^4 z^2\,})\, ,
\ee
and expand as
\begin{equation}
 {\cal L}_{\rm BI}= - t
 + \frac{1}{2} g^2 \left(t^2+z^2\right)
 -\frac{1}{2} g^4 t \left(t^2+z^2\right)
 +\frac{1}{8} g^6 \left(t^2+z^2\right) \left(5 t^2+z^2\right)
  -\frac{1}{8} g^8 t \left(t^2+z^2\right) \left(7 t^2+3 z^2\right)+\cdots\,.
\end{equation}

We continue the discussion of the BI case soon, but first we will discuss a distinct non-linear deformation of electromagnetism. While superficially complicated, this next deformation is, in fact,  much easier  to generate from pure Maxwell electrodynamics.  Indeed we will see a tradeoff between the relative simplicity of the deformed action  in the BI case and the complicated initial deformation source required to generate it and the relative simplicity of the initial deformation source which results in the superficially complicated action we will now present.

\subsection{Bossard-Nicolai Model}
\label{BNLagrangianSection} 
With the same  variables $t$ and $z$, one can write the following NGZ-consistent Lagrangian 
\begin{multline}
\label{BNaction}
{\cal L}_{\rm BN} = -t + \frac{1}{2} g^2 \left(t^2+z^2\right)  -\frac{1}{2} g^4\, t\, \left(t^2+z^2\right) + \frac{1}{4} g^6 \left(t^2+z^2\right) \left(3 t^2+z^2\right)
-\frac{1}{8} g^8\, t \, \left(t^2+z^2\right) \left(11 t^2+7 z^2\right) \\  
+ \frac{1}{32} g^{10} \left(t^2+z^2\right) \left(  91  t^4+  86 t^2 z^2+11 z^4\right) 
-\frac{1}{8} g^{12} \, t\,  \left(t^2+z^2\right) \left(51 t^4+64 t^2 z^2+17 z^4\right)\\
+\frac{1}{64} g^{14} \left(t^2+z^2\right) \left(969 t^6+1517 t^4 z^2+623 t^2 z^4+43 z^6\right) + \cdots \, .
\end{multline}
One simply keeps adding terms necessary so as to maintain the consistency 
\eqn{consistencyEqn} order by order, specifically via a procedure we will discuss in~\sect{BNdualitySection}.  Unlike the Born-Infeld action,  we do not know if this has a closed-form expression.
 Note that this Lagrangian differs from ${\cal L}_{\rm BI}$  starting at ${\cal O}(g^6)$.

It is not difficult to verify that \eqn{consistencyEqn} is maintained order by order.  Using,
$\tilde G= 2{\partial L\over \partial F}= ( \partial_t L) F+ ( \partial_z L) \tilde F$ and $G=  - (\partial_t L)  \tilde F+ (\partial_z L) F$,
we can rewrite the NGZ identity as,
\be
\label{lagrangeNGZ}
\big( (\partial_t L)^2-(\partial_z L)^2-1\big)\,z -\big(2 \, (\partial_z L) (\partial_t L)\big)\, t=0
\ee

Although the explicit Lagrangian \eqn{BNaction} is not provided in ref.~\cite{BN}, it is indeed the non-linear deformation of classical electrodynamics that is produced\footnote{Strictly speaking ref.~\cite{BN} presents  this model with negative $g^2$ so as to generate a positive Hamiltonian, as discussed in~\app{HamiltonianAppendix}.}  order by order as we will describe shortly.

\section{Twisted Self-duality Constraints}
\label{twistedSelfDualConstraintSection}
While the duality constraints are readily checked in the two above examples, BI and BN, note that, by hand, we forced a functional form of $G$ in terms of $F$ through eq.~(\ref{covConstraint}).  The very act of doing so, prioritizing the primacy of one over the other, makes the duality between $F$ and $G$ no longer manifest.    We can avoid this by introducing what has been called a ``twisted self-duality'' constraint -- a constraint that guarantees that only one vector field from the duality doublet will ever be independent, but without establishing priority for one over the other.  
This constraint generalizes the equation~(\ref{covConstraint}), in that it can be considered more fundamental than
the Lagrangian ${\cal L}$ which it, in fact, determines. 
The symmetry between $F$ and $G$ will only be broken by the solution to this constraint.

\subsection{Schr\"odinger's BI Solution} 

In the Born-Infeld example,  such a constraint was first found by Schr\"odinger in 1935~\cite{Schrodinger}. To describe Schr\"odinger's construction in the form given in~\cite{Gaillard:1997rt} it is useful  to consider  the duality symmetry in a complex basis where
 \be
 \label{defT}
T=F-iG\,, \hskip 1cm   \hskip 1cm  \,  T^*= F+iG\, ,
\ee
and the $U(1)$ duality symmetry is
\begin{eqnarray}
\delta \left(
                      \begin{array}{cc}
                  F-iG \\
                  F+iG \\
                      \end{array}
                    \right)\ =\left(
                      \begin{array}{cc}
                      iB&  0 \\
                       0 & -iB \\
                      \end{array}
                    \right)  \left(
                      \begin{array}{cc}
                   F-iG \\
                      F+iG \\
                      \end{array}
                    \right) \, .
\end{eqnarray}
 Schr\"odinger suggested the following  exact  duality covariant cubic  self-duality constraint
\be
T_{\mu\nu}  (T\tilde T  ) - \tilde T_{\mu\nu} T^2 = {g^2\over 8} \tilde T^*_{\mu\nu}  (T\tilde T )^2 \ .
\label{exact}\ee
It is straightforward to verify that, if this constraint is solved perturbatively, one finds the unique Born-Infeld solution of the NGZ identity
\be
T\tilde T^*=F\tilde F + G\tilde G=0 \ .
\label{NGZcpx}
\ee
And, even better, there is an action which is manifestly duality invariant~\cite{Schrodinger,Gaillard:1997rt},
\be
\label{SchLagrange}
{\cal L}_{\rm Sch}(T)= 4 {T^2\over (T\tilde T)}\, , \qquad {\cal L}_{\rm Sch}= - {\cal L}_{\rm Sch}^* \ .
\ee
This fascinating Lagrangian is a ratio of two duality invariants
\bea
T^2&=& (F-iG)^2= F^2-2iFG - G^2 \, , \\
T\tilde T&=& (F-iG) (\tilde F-i\tilde G)\\
&=& F\tilde F -2i F\tilde G- G\tilde G \nn\,.
\eea
The cubic constraint (\ref{exact}) is equivalent to the requirement that the  derivative of the Schr\"odinger action ${\cal L}_{\rm Sch}(T)$  over $T$ defines the conjugate $\tilde T^*$:
\be
\label{exactSch}
 \tilde T^*_{\mu\nu} \equiv g^{-2}  \frac{\partial {\cal L}_{\rm Sch}}{ \partial T^{\mu\nu}}  \ .
\ee
It follows that
\be
\label{exactSch1}
{\partial {\cal L}_{\rm Sch}
\over \partial T^{\mu\nu}}=  8\Big (T_{\mu\nu} {1\over  (T\tilde T  )} - \tilde T_{\mu\nu} {T^2  \over (T\tilde T )^{2}}\Big )= {g^2} \tilde T^*_{\mu\nu} \ .
\ee
Contraction with $T^{\mu\nu}$ demonstrates that (\ref{NGZcpx}) holds.

To make contact with the supergravity formalism and the discussion in Appendix~\ref{sugraGenApp}, we  introduce self-dual notation, 
\be
\label{selfDualT}
T^\pm= \textstyle{ \frac{1}{ 2}}(T\pm i \tilde T) 
\ee
such that   $T^+_{\mu\nu} T^{-\mu\nu}=0$ and
\be
T^*=(T^{*})^+ + (T^{*})^- \qquad  (T^{*})^\pm= \textstyle{\frac{1}{ 2}}(T^*\pm i \tilde T^*) \,. \ee 

Recalling that $(\tilde T)^2=-T^2$, we have 
\be
\label{Tident}
T^2 - i(T\tilde T  )=T(T-i \tilde T) = 2 TT^-= 2 (T^-)^2\,.
\ee
We can now rewrite the cubic  self-duality constraint \eqn{exactSch} as
\be
  T_{\mu\nu}^+ (T^-)^2 +  {g^2\over 16} (T^{*})^+_{\mu\nu}  (T\tilde T )^2=0\,,
\label{cubiccomb} 
\ee  or
\be
  T_{\mu\nu}^+ (T^-)^2 -  {g^2\over 16} T^{*+}_{\mu\nu}  \Big ((T^+)^2- (T^-)^2 \Big )^2=0\,,
\label{cubiccomb1} 
\ee
and the NGZ identity (\ref{consistencyEqn}) is
\be
T^{*+}T^+- T^{*-}T^-=0\,.
\label{NGZsdcpx}
\ee
This formulation of the NGZ identity will be useful in later sections.

\subsection{Maxwell Case}
Note that in the Maxwell case with $g=0$ there is a particularly simple duality covariant linear twisted self-duality constraint $G=\tilde F$ and $F=-\tilde G$, which in self-dual notation is
\be
\label{maxwellduality}
T^{+}= F^+- i  G^+=0 
\ee
and does indeed follow from the $g^2\to0$ limit of eq.~(\ref{cubiccomb1}). The conjugate of (\ref{maxwellduality}) is $ (T^{+})^*= F^-+ i  G^-=0$. It should be noted, however, that \eqn{cubiccomb1} cannot be interpreted as a local perturbative deformation of (\ref{maxwellduality}).

\subsection{BN Case}
\label{BNdualitySection}
In contrast, the model in \eqn{BNaction} which is consistent with NGZ identity satisfies a local deformation 
of (\ref{maxwellduality}), in which the right-hand side is modified as
\be
\label{BNdualityT}
T^+_{\mu\nu}= {g^2\over 16}  T^{*+}_{\mu\nu} (T^-)^2  \,.
\ee
Using eqns.~(\ref{covConstraint}), (\ref{defT}), (\ref{selfDualT}), and 
\bea
  G^+ &=& {\textstyle \frac{1}{2}} ( G + i \tilde G)\nonumber \\
          &=& {\textstyle \frac{1}{2}} ( F + i \tilde F) (\partial_z {\cal L} + i \partial_t {\cal L})\nonumber\\
           &=& F^+ (\partial_z {\cal L} + i \partial_t {\cal L})\,,
\eea
we can translate \eqn{BNdualityT} back into constraints on derivatives of the action, 
\begin{equation}
\label{BNduality}
0=(1 + \partial_t {\cal L} -i \, \partial_z {\cal L})  
 - { g^2 \over   8}  (t-i z) (1- \partial_t {\cal L} -i \, \partial_z {\cal L} )^2   (1-  \partial_t {\cal L} + i \, \partial_z {\cal L} )\,.
\end{equation}
Foreshadowing slightly -- requiring analyticity of ${\cal L}$ for small values of $F$ -- one may introduce an ansatz in terms of monomials in $g^2$, $t=F^2/4$, and $z=F \tilde F/4$, 
\be
\label{lagrangeAnsatz}
{\cal L}=\Big(g^{-2} \!\!\!\!\!  \sum_{m=0, p=0} \!\!\!\!\!  
  g^{2 (p + 2 m)} c_{(p, 2 m)} t^p z^{2 m}\Big)- c_{(0,0)} g^{-2} \, ,
\ee
and solve \eqn{BNduality} algebraically, order by order in $g^2$, fixing the constant coefficients $c_{(i,j)}$.  Doing so results in a Lagrangian which satisfies the NGZ equation, and reproduces \eqn{BNaction}. 

 Indeed, as we will see, the covariant procedure proposed in ref.~\cite{BN} is to modify the linear twisted self-duality constraint to a  non-linear duality constraint by the introduction of a single deformation (or counterterm) as we just did to go from \eqn{maxwellduality} to \eqn{BNdualityT}.  It so happens that in the cases studied in ref.~\cite{BN},  as with \eqn{BNdualityT}, a single such deformation was sufficient.   We can see already, given the cubic nature of the BI constraint, that in general we will require a procedure which introduces an infinite number of such deformations to the linear twisted self-duality constraint.    Indeed  the non-covariant procedure of 
 Floreanini, Jackiw, 
 Henneaux and Teitelboim~\cite{Floreanini:1987as, Henneaux:1988gg}, discussed in ref.~\cite{BN} has the potential to allow an infinite amount of information.  Ref.~\cite{BN} seemed to constrain its constants of integration to explicitly reproduce the covariant procedure described above and more generally in~\sect{BNproc}.  This need not be so.   The generalization of the covariant procedure discussed in~\sect{genProcSection} can be arrived at non-covariantly by allowing arbitrary constants of integration that satisfy  the relevant NGZ relation. We have in fact verified that the Born-Infeld Hamiltonian can be obtained in this approach, see Appendix~\ref{HamiltonianAppendix}.
 
\section{Bossard-Nicolai (BN) proposal}
\label{BNprop}

  We start by explicitly providing an algorithm for the covariant procedure introduced in ref.~\cite{BN}.  We  subsequently review the provided supporting examples.

\subsection{Covariant BN procedure}
\label{BNproc}

Bossard and Nicolai posit~\cite{BN}  the existence of procedures which would allow the deformation of all classically duality invariant theories, including ${\cal N}=8$ supergravity.  
This proposal was worked out on three examples in ref.~\cite{BN}, and here we reconstruct the covariant procedure in detail.   

 A convenient language for extended supergravities  comes from the fact that any candidate counterterm would depend on the graviphoton\footnote{See eq. (\ref{GP}) for definition of this particular combination of $F$ and $G$ and scalars for supergravities with scalars in the $G/H$ coset space.}. More specifically the counterterm would depend on the conjugate self-dual field strength ${\overline T}{}^{+AB}$ and the anti-self-dual field strength $T^{-}_{AB}$. 
In the $G/H$ coset space, $AB$  are the indices of the antisymmetric representation of the group $H$. 
 For example, for ${\cal N}=8$ supergravity these would be $SU(8)$ indices (in the 28-dimensional representation) and  $G/H$ is $E_{7(7)}/SU(8)$.  For $U(1)$ the deformation source depends on $T^{*+}$ and $T^{-}$.  In this procedure, as with the generalized procedure we present in \sect{genProc}, we will include the $H$-symmetry indices.  The same procedures work for $U(1)$ with the indices elided. 

One starts with  an initial action $S_{\rm init}$ with a conserved duality current and a manifestly duality-invariant  counterterm, or deformation, $\Delta S$.  
It is assumed that $\Delta S$ can be expressed as a manifestly duality invariant function of $F$ and $G$ or, equivalently, 
on  $\overline{T}{}^{+AB}$ and $T^{-}_{AB}$. Classically $T^+_{AB}=0$ is the linearized twisted self-duality constraint, which we will be deforming.  The goal is to construct 
a Lagrangian ${\cal L}_{\rm final}$ that incorporates the counterterm/deformation yet still conserves the duality current.  For the general case this means satisfying NGZ identity given in \eqns{const}{deltaS},  and the simpler (\ref{consistencyEqn}) for $U(1) $.  Of course, one should also require that it possesses the field content and other relevant symmetries of
  $S_{\rm init}$.  The construction proceeds as follows:

\begin{enumerate}

\item Take the variation of the counterterm with respect to the field-strength,  and express as a function of $T^-$, and $\overline{T}{}^+$ which we will call the {\it initial deformation source} ${\cal I}^{(1)}$ 
\be
\frac{\delta \Delta S}{\delta \overline{T}{}^{+AB}} \to \frac{\delta {\cal I}^{(1) }(T^-_{AB}, \overline{T}{}^{+ AB})}{ \delta \overline{T}{}^{+AB}}
\ee

\item  Constrain the self-dual field strength to the variation of this initial source: 
\be
\label{def}
T^+_{AB}=\frac{\delta {\cal I}^{(1) }(T^-_{AB}, \overline{T}{}^{+AB})}{ \delta \overline{T}{}^{+AB}}
\ee
This is a modification of the linear twisted self-duality constraint $T^{+AB}$=0. \footnote{When ${\cal I}^{(1) }$ has only terms quadratic in $T$ (as in $U(1)$ and  the toy model ${\cal N}=8$ examples of sec. 2 in ref.~\cite{BN}), the right-hand side of \eqn{def} remains linear in $T$ so the deformation of the linear constraint remains linear.}

\item Translate \eqn{def} to a differential constraint  on $S_{\rm final}$, c.f. \sect{BNdualitySection} for the $U(1)$ case.

\item Introduce an ansatz for ${\cal L}_{\rm final}$ in terms of the Lorentz invariants, c.f. \eqn{lagrangeAnsatz}, again for the $U(1)$ case.  This will be more complicated, of course, for the generic case.

\item Solve for the ansatz order by order in the coupling constant, at each step verifying the consistency of the relevant NGZ relation, the presence of additional desired symmetries of the target Lagrangian and enlarging the ansatz if one runs into an inconsistency.
\end{enumerate}

In contrast to ref.~\cite{BN} we do not call ${\cal I}^{(1)}$ the ``initial deformation.''  As we will see in the generalized procedure in order to even recover the Born-Infeld action we will need to include an infinite number of terms to modify the covariant twisted self-duality constraint.  One can integrate those infinite deformations to achieve a final ${\cal I}_{\rm BI}$, but this will not be the final deformation of the action ${\cal L}_{\rm Max}-{\cal L}_{\rm BI}$, rather it is simply the complete source of the deformations to the linear twisted self-duality constraint required to generate the BI deformation of the action through the generalized procedure.  For consistency, then, we refer to ${\cal I}^{(1)}$ as the initial deformation source.

\subsection{Three BN examples}

Two examples of the deformation of the linear twisted self-duality condition discussed in ref.~\cite{BN} relate to Maxwell electrodynamics and one to a toy model of ${\cal N}=8$ supergravity. 

The first example, from sec. 2 of ref.~\cite{BN}, is a Maxwell deformation analogous to an  ${\cal N}=8$ supergravity counterterm.  The deformation is quadratic in $F$, with derivatives of the Maxwell field, ${\cal I}^{(1)}\sim C^2 (dF)^2$.  The dependence on derivatives necessitates  the  following deformed twisted self-duality constraint~\cite{Chemissany:2006qd}
\be
{\delta \over \delta F(y)} \int d^4x (\tilde GB G+ \tilde F BF)=0\,.
\label{der1}\ee
In this case { $G$ is linear in $F$  and the action remains quadratic in $F$}. The reconstruction is based on NGZ identity in the form $S= \frac{1}{4} F\tilde G$ which is valid only for the actions quadratic in $F$ when $S_{\rm inv}=0$ in eqs. (\ref{sinv}) and (\ref{reconstr}).  As the result of the deformation (\ref{def}) the reconstructed action $S(F)$ has some non-polynomial non-local terms required to complete the deformation in the action. 
This example, however, has linear duality since $G$ remains a linear function of $F$ even with the deformation caused by ${\cal I}^{(1)}\sim C^2 (dF)^2$.

A closely related example in sec. 2 is a  toy model of an ${\cal N}=8$ supergravity deformation caused by the part of the three-loop counterterm which is  quadratic in $F$ and quadratic in Weyl curvatures.  The quartic in $F$ terms  $(\partial F)^4$ present in the ${\cal N}=8$  three-loop counterterm,   $C^4+(\partial F)^4 + C^2(\partial F)^2+ \cdots$, are { not} taken into account in this example. This example, therefore is also of the type given in eqs. (\ref{sinv}) and (\ref{reconstr}) where $S= \frac{1}{4} \int F^\Lambda \tilde G_\Lambda+S_{\rm inv} $ and ${\delta S_{\rm inv}\over \delta F}=0$. In the toy model $\tilde G$ remains a linear function of $F$, in absence of contribution to the right-hand side of eq.  (\ref{reconstr})  from ${\delta S_{\rm inv}\over \delta F}=0$,  and therefore the  linear duality of the classical action is preserved by deformation.  Note that in the case of linear duality the action is easily reconstructed, all dependence on vectors is in $S_{\rm vect}= \frac{1}{4} \int F^\Lambda \tilde G_\Lambda$ and it satisfies NGZ identity  as explained in (\ref{linearvar}).
Thus, this example also does not immediately shed light on cases of non-linear duality when the vector dependent part of $S_{\rm inv}$ is present and contains $(\partial F)^4$ terms which require the presence of all increasing powers of $F$.

  In both examples of sec. 2  in ref.~\cite{BN} a  Lorentz covariant single term deformation of the undeformed constraint is employed as shown in \eqn{def}.

The third example is the deformation we discussed as the BN model earlier in \sect{BNdualitySection}.
Without derivatives in $F$, the manifestly $U(1)$ invariant `initial' deformation source, quartic in $F$ , is used in the Lorentz covariant cubic deformation of the linear constraint (\ref{def}), and its equivalent Hamiltonian formulation.    The proposed procedure is equivalent to the one worked out earlier: introduce the initial source, and then solve the twisted self-duality constraint  for a Lagrangian order by order by introducing an ansatz polynomial in the available Lorentz invariants.

Any procedure must require that the deformed action, reconstructed using the deformed twisted self-duality constraint (\ref{def}), satisfies the relevant NGZ constraints (\ref{consistencyEqn}).  All examples considered in~\cite{BN} have the nice property that the only input into the right-hand side of (\ref{def}) is a term ${\cal I}^{(1)}$ quadratic or quartic in field strengths,  and they  indeed satisfy the relevant NGZ constraints: (\ref{der1}) in the case with derivatives and (\ref{consistencyEqn}) in models without derivatives on $F$. 
No allowance is made, however, for  cases when the solution of eq. (\ref{def}) is inconsistent with direct higher-loop calculations,
as neither of the examples indicated the need for such a possibility.

We will see that the Born-Infeld model requires the presence of an infinite set of deformations of the linear constraint (\ref{maxwellduality}).  Instead of \eqn{def}, we will find that a general procedure will impose,
\be
\label{defn}
T^+_{AB}=\frac{\delta {\cal I}^{(1)}}{ \delta {\overline T}{}^{+AB}}+\cdots+\frac{\delta {\cal I}^{(n)}}{ \delta {\overline T}{}^{+AB}} +\cdots=
\frac{\delta {\cal I}( T^{-}_{AB}, {\overline T}{}^{+AB}, g)}{\delta {\overline T}{}^{+AB}} \,,
\ee
where the various terms need not  be related to the initial ${\cal I}^{(1)}$.
In the following section we present a procedure that successfully reproduces the Born-Infield deformation.

\section{Generalized Covariant Procedure}
\label{genProcSection}
First we present the procedure that we use to recover the Born-Infeld deformation in the BN framework, and see that it does, indeed, require an infinite number of modifications to the linear twisted self-duality constraint.   Learning from this example we modify the procedure of \sect{BNproc} so as to handle the more general case.


\subsection{Finding the Born-Infeld Deformation}
\label{biDeformProc}

We can begin by introducing an ansatz for the deformation source ${\cal I}( T^-,T^{*+}, g)$ in terms of a series expansion,  i.e.
\be
\label{biAnsatz}
T^+_{\mu \nu} =  {g^2\over 16}  {T^*}{}^{+}_{\mu\nu} (T^-)^2  \Big[\, 1 + \sum_{n=0} d_n \Big( {\textstyle \frac{1}{4}}\, g^4  ({ T^*}{}^{+})^2(T^-)^2 \Big)^{n} \,  \Big],
\ee
where $d_{n}$ are the real parameters to be constrained so as to reproduce the Born-Infeld deformation.  Since we are looking to reproduce the BI Lagrangian,  and we know it ahead of time, we may simply set ${\cal L}$ to \eqn{BILag}. It is not difficult to check (by multiplying with ${\overline T}{}^{+}$ and subtracting from the result the product between $T^-$ and the conjugate of (\ref{biAnsatz})) that there exist solutions obeying the NGZ identity (\ref{NGZsdcpx}).

As in \sect{BNdualitySection}, we can  translate \eqn{biAnsatz}  into constraints on derivatives of the BI action using  $G^+ = F^+ (\partial_z {\cal L} + i \partial_t {\cal L})$,
\begin{multline}
\label{BIdualityAns}
0=(1 + \partial_t {\cal L} -i \, \partial_z {\cal L})+ { g^2 \over   8}  (t-i z) (1- \partial_t {\cal L} -i \, \partial_z {\cal L} )^2   (1-  \partial_t {\cal L} + i \, \partial_z {\cal L} )  \Big[ 1+\\
 \sum_{n=0} d_{n}  \Big( g^4   (t-i z) (1- \partial_t {\cal L} -i \, \partial_z {\cal L} )^2  (t+i z) (1-  \partial_t {\cal L} + i \, \partial_z {\cal L} )^2 \Big)^{n}\, \Big]  \,.
\end{multline}

We expand in a series of the coupling constant and solve for $d_{n}$ order by order.   We indeed find an infinite series which we can express as a generalized hypergeometric function so the BI twisted self-duality constraint can be given, 
\be
\label{biNewduality}
T^+_{\mu\nu}= \textstyle{\frac{1}{16}}\,g^2\,{\overline T}{}^{+}_{\mu\nu} \, 
  (T^-)^2 \,\,{}_{3}{\rm F}_{2}\big(\textstyle{\frac{1}{2}, \frac{3}{4}, \frac{5}{4}};\textstyle{\frac{4}{3}, \frac{5}{3}};
\textstyle{-\frac{1}{27}}\,{g^4 \,({\overline T}{}^{+})^2\,(T^-)^2 } \,\big) \,.
\ee
Writing eq.~(\ref{biNewduality}) as
\be
T_{\mu\nu}^+  = \frac{\delta {\cal I}(T^-, \bar T^+, g)}{\delta \bar T^{+}_{\mu\nu}}
\label{BNBI}\ee
we find that 
the required deformation source takes the following form
\be
{\cal I}(T^-, \overline T^+, g)={6\over g^2}
\Big(1-
{}_3F_2(\textstyle{-\frac{1}{2}, -\frac{1}{4}, \frac{1}{4}};\textstyle{\frac{1}{3}, \frac{2}{3}};
\textstyle{-\frac{1}{27}}\,{g^4 \,({\overline T}{}^{+})^2\,(T^-)^2 })
\Big)
\label{calI}\ee
The procedure then for deforming to BI is to modify  \eqn{maxwellduality} to \eqn{biNewduality} and then to introduce an ansatz for the Lagrangian  to be solved for order by order. The resulting Lagrangian should be analytic for small values of the field strength.

We have therefore constructed (\ref{calI}) a deformation source ${\cal I}(T^-, \bar T^+, g)$  which, like 
Schr\"odinger's action  ${\cal L}_{\rm Sch}(T)= 4 {T^2\over (T\tilde T)}$ via \eqn{exactSch}, yields a twisted self-duality constraint  whose solution is the Born-Infeld action. 
The differences between the two expressions are striking; moreover, while both are duality invariant, 
their natural variables and, consequently, the resulting deformed twisted self-duality constraints, 
(\ref{exactSch1}) and (\ref{biNewduality}), are different. 
This opens the possibility that there may exist other deformations, different from them, which nevertheless generate the same duality-invariant action. It would be interesting to explore this possibility as well as the relation between these actions.


\subsection{Generalized Covariant Procedure}
\label{genProc}

Thus, to reproduce a sufficiently general action with a conserved duality current, we must allow the counterterm to be a general function of the coupling constant and duality invariants which is analytic for small values of fields.
As before, we present this discussion in terms of graviphoton field strengths (see \app{sugraGenApp}), but the $U(1)$ examples follow by simply dropping the indices.

 We start with a duality conserving initial action $S_{\rm init}$, and a duality-invariant  counterterm, or deformation, $\Delta S$.  We assume, as BN, that $\Delta S$ can be expressed as a function the conjugate self-dual field-strength $\overline{T}{}^{+AB}$.  We wish to arrive at a Lagrangian ${\cal L}_{\rm final}$ that incorporates the counterterm yet still conserves the duality current.  We proceed as follows:
\begin{enumerate}

\item Take the variation of the counterterm with respect to the field-strength,  and express as a  function of $T^-$, and $\overline{T}{}^+$, 
\be 
\frac{\delta \Delta S}{\delta {\overline T}{}^{+AB}} \to 
\frac{\delta {\cal I}(T^-_{AB}, {\overline T}{}^{+ AB}, g)}{ \delta {\overline T}{}^{+AB}}
\ee

\item Introduce an ansatz for the deformation source ${\cal I}( T^{-}_{AB}, {\overline T}{}^{+AB}, g)$. In general, this may be taken to depend on all possible duality invariants\footnote{In the case of the non-linear 
$U(1)$ duality we assumed that ${\cal I}$ is an analytic function of $g^4  ({\overline T}{}^{+})^2(T^-)^2$.  There is, 
however, in  more general theories no reason to forbid higher-order 
counterterms.  In other words, if we have to worry about adding counterterms, we might as well worry about adding all counterterms allowed by the known symmetries. E.g. for ${\cal N}=8$ supergravity we should at least include in the ansatz { all} $E_{7(7)}$ invariants.}.

\item Constrain the self-dual field strength to this variation: 
\be
\label{defGen}
T^+_{AB}=\frac{\delta {\cal I}( T^{-}_{AB}, {\overline T}{}^{+AB}, g)}{ \delta {\overline T}{}^{+AB}}
\ee

\item Translate \eqn{defGen} to a differential constraint  on ${\cal L}_{\rm final}$, c.f. \sect{biDeformProc} for the $U(1)$ case.  The differential constraint in general is more complicated, see (\ref{const}), 
(\ref{deltaS}).

\item Introduce an ansatz for ${\cal L}_{\rm final}$  which is analytic around the origin in terms of the Lorentz invariants.  For the case of $U(1)$, again, this was not so difficult (\eqn{lagrangeAnsatz}), but in general this is unknown and can depend on other fields (e.g. scalars) in non-trivial ways.

\item Solve for { both} the ${\cal I}$ ansatz  parameters, as well as the 
Lagrangian ansatz parameters, order by order in the coupling constant, enforcing 
the consistency of the relevant NGZ  consistency equation (in $U(1)$ case any of the eqs. (\ref{consistencyEqn}), (\ref{NGZsdcpx}) or (\ref{lagrangeNGZ})),  and additional desired 
symmetries of the target Lagrangian, enlarging the ansatz if one runs into inconsistency.
\end{enumerate}

The procedure given in \sect{BNproc} is recovered by restricting to the lowest order term 
in the small $g$ expansion of ${\cal I}$. 
We also see that, at least for deformations of Maxwell's theory, there are an infinite number 
of classical solutions recoverable by this procedure, consistent with the findings of 
ref.~\cite{Gibbons:1995cv, Gaillard:1997rt} where it was shown that the NGZ identity (\ref{consistencyEqn}) has 
infinitely many solutions.   

There exists the possibility that the counterterms generated by iterating on some first counterterm 
${\cal I}^{(1)}$ differ at some loop level from counterterms discovered  by explicit calculation. 
Unlike the original procedure, if the difference is a duality invariant, our strategy can accommodate it by a suitable modification of 
${\delta {\cal I}}(T^-_{AB},{\overline T}^{+AB}, g)$. In the supersymmetric context discussed in the next section this allows for complete supersymmetric invariants to be independently included starting at some loop order higher than the one at which the first counterterm appears.

It is important to note that in the $U(1)$ case without derivatives and scalars, a hermitian deformation and manifestly $U(1)$ invariant deformation ${\cal I}(T^-,  T^{*+}, g)$ guarantees that the NGZ equation is satisfied. Indeed, using (\ref{defGen}) it is easy to see that 
\be
T^{*+}\frac{\delta {\cal I}(T^-,  T^{*+}, g)}{\delta  T^{*+}} - T^- \frac{\delta {\cal I}(T^-,  T^{*+}, g)}{\delta  T^{-}}=T^{*+}T^+- T^{*-}T^-=0\,.
\label{NGZsdcpx1}
\ee
This was manifestly the case for the deformation ansatz for any real choice of $d_n$ in \eqn{biAnsatz}.  This is in contrast to the NGZ equations relevant for supergravity as we will discuss in \app{sugraGenApp}.

\section{Nonlinear $U(1)$ duality and supersymmetry}
\label{SUSY}

The NGZ condition for $U(1)$ duality invariance (\ref{consistencyEqn})  has infinitely many solutions which are analytic 
for sufficiently small field strength~\cite{Gibbons:1995cv, Gaillard:1997rt}. As we saw in earlier sections, the BN 
deformed self-duality constraint selects one such solution. In the case of Maxwell's theory deformed
by a quartic interaction the resulting action, while self-dual, differs from the Born-Infeld action 
starting from the sixth order terms. 
By allowing higher order deformations it is possible to accommodate the Born-Infeld action in the 
deformed self-duality framework. This generalization of the BN proposal, while necessary to include 
known examples of nonlinear duality in this framework, also leads to an apparent loss of predictive 
power by allowing us to freely deform the action order by order in perturbation theory.
Assuming that we did not know of the Born-Infeld action, we would like to find a physical principle
that singles it out of this infinite family of duality-invariant actions. More generally, we would like to 
find a principle that selects physically-relevant actions. 

Since Maxwell theory can be supersymmetrized up to maximal supersymmetry, it is natural to 
require that this feature survives the nonlinear extension. A similar requirement arises 
naturally if one considers applying the twisted self-duality ideas to (maximal) supergravity. 
We will therefore explore the conditions under which twisted self-duality is compatible with 
minimal and extended supersymmetry. In this discussion of supersymmetry and self-duality we follow mostly the work by Kuzenko and Theisen \cite{Kuzenko:2000uh} and Ketov  \cite{Ket2}.

\subsection{${\cal N}=1$ supersymmetric  nonlinear electrodynamics}

Models with nonlinear $U(1)$ duality and ${\cal N}=1$ supersymmetry are constructible in 
superspace, see \cite{Cecotti:1986gb} and \cite{ Kuzenko:2000uh,Aschieri:2008ns}. 
The action is constructed from the standard (anti)chiral 
field-strength superfields
\be
W_\a = -\frac{1}{4}\, {\overline D}{}^2 D_\a \, V \, , \qquad \quad 
{\overline W}_\ad = -\frac{1}{4}\,D^2 {\overline D}_\ad \, V  \, ,
\label{w-bar-w}
\ee
defined in terms of  a real unconstrained prepotential $V$.  The Bianchi 
identities
\be
D^\a \, W_\a  \, = \,  {\overline D}_\ad \, {\overline W}{}^\ad
\label{n=1bi}
\ee
are automatically satisfied.  Similarly to the bosonic case, the dual (anti)chiral field strengths, 
${\overline M}_\ad$ and $M_\a$, are defined from the action $S[W,{\overline W}]$
as follows
\be
{\rm i}\,M_\a \,[W]\equiv 2\, \frac{\d }{\d W^\a}\,S[W,{\overline W}]
 \, , \qquad \quad
- {\rm i}\,{\overline M}{}^\ad \,[W]\equiv 2\, 
\frac{\d }{\d {\overline W}_\ad}\, S[W,{\overline W}]  \, .
\label{n=1vd}
\ee
The equations of motion for the vector multiplet 
may be expressed in terms of $M$ and ${\overline M}$ as
\be
D^\a \, M_\a  \, = \,  {\overline D}_\ad \, {\overline M}{}^\ad \,.
\label{n=1em}
\ee
The superysmmetric generalization 
of the NGZ relation  requires that
\be 
{\rm Im} \int d^4 x d^2\theta \, 
\Big( W^\a W_\a  \, + \,  M^\a M_\a \Big)  \, = \, 0 \, .
\label{n=1dualeq}
\ee
One may understand the structure of this relation by recalling that the bosonic NGZ relation
is quadratic in field strengths in addition to being invariant under the infinitesimal duality rotation 
\be
\delta F = \lambda G \, ,
\qquad\qquad\qquad
\delta G = -\lambda F \ .
\ee

The Bianchi identities (\ref{n=1bi}) and the equations of motion (\ref{n=1em})  are 
therefore invariant under a similar transformation acting on $W$ and $M$. 
Moreover, the supersymmetric NGZ identity \eqn{n=1dualeq} is also invariant under this transformation.
It is worth noting that this equation reduces to the bosonic NGZ relation \eqn{consistencyEqn} upon setting the 
fermion and auxiliary fields to zero.

The ${\cal N}=1$ Maxwell theory is a solution of \eqn{n=1dualeq}. To construct interacting theories which solve
the supersymmetric NGZ relation one may start, following ref.~\cite{Kuzenko:2000uh}, with a general 
action 
\be
{\cal S}  \, = \,  \frac{1}{4}\int {\rm d}^6z \, W^2 +
\frac{1}{4}\int {\rm d}^6{\bar z} \,{\bar  W}^2 
+  \frac{1}{4}\, \int {\rm d}^8z \, W^2\,{\overline W}{}^2  \,
\L \Big( \frac{1}{8} D^2\,W^2 \, ,\, \frac{1}{8}
{\bar D}^2\, {\overline W}{}^2 \Big)
\label{gendualaction}
\ee
parametrized by the real analytic function of one complex variable $\Lambda(u, {\bar u})$. Constructing 
the dual super-field strengths (\ref{n=1vd}) it is not difficult to find that the NGZ constraint requires that 
$\Lambda$ be a solution of 
\be
{\rm Im}\;  \left\{ \partial_u  (u \, \L) 
- \bar{u}\, 
\left( \partial_u  (u \, \L ) \right)^2 \right\} = 0 \, .
\label{GZ4'}
\ee
This partial differential equation has infinitely many solutions, parametrized {e.g.} by the 
coefficients of the terms $(u{\bar u})^n$ with $n\ge 2$ in the expansion around $u=0$ (as well as 
the coefficient of $u{\bar u}^2$). This 
freedom is sufficient to accommodate all the solutions of the bosonic deformed  
self-duality constraints discussed in earlier sections. 

Indeed, taking the integral over the fermionic superspace coordinates, and setting the gauginos and auxiliary 
fields\footnote{This is consistent, as the auxiliary fields alway appear squared after all 
supersymmetric covariant derivatives are evaluated in \eqn{gendualaction}.}  to zero, we find 
\be
L   \, = \,  -\hf ( {\bf u} + {\bar {\bf u}} )  \, + \, 
{\bf u}  {\bar {\bf u}}\, \L({\bf u}, {\bar {\bf u}}) \, , \qquad
{\bf u}  \,  \equiv  \,  \frac{1}{8} D^2 W^2 \big|_{\q = 0, D=0,\psi=0}  \, = \,  
\frac{1}{4}F^2+\frac{i}{4}F{\tilde F}\equiv\o  \, .
\ee
It is not difficult to see that it is possible to choose functions $\Lambda$ such that this Lagrangian 
reproduces the two solutions discussed explicitly in \sect{dualityEM}. The choice of $\Lambda$ for the 
Born-Infeld Lagrangian, \sect{BILagrangianSection}, is well-known~\cite{Kuzenko:2000uh}
\bea
L_{\rm BI} &=& \frac{1}{g^2} \Big\{\;
1 - \sqrt{- \det (\eta_{ab} + g F_{ab} )} \;
\Big\} 
=
 \frac{1}{g^2} \Big[ \;
1 - \sqrt{1 + g^2 (\o + \bar \o )  
+{1 \over 4}g^4 (\o - \bar \o )^2 } \;
\Big] \, , \non \\
\L_{\rm BI}  &=&
\frac{g^2 }{ 1 + 
\hf g^2(\o + \bar \o )
+ \sqrt{1 + g^2 (\o + \bar \o )  
+\frac{1}{4}g^4 (\o - \bar \o )^2 }} \, .
\label{bi-lag}
\eea 
The Lagrangian obtained with the BN deformation, \sect{BNLagrangianSection}, may be expressed 
in terms of $\o$ as 
\bea
L&=&-\frac{1}{2}(\o+\bar\o) + \frac{g^2}{2}\o\bar\o 
     -\frac{g^4}{4}\o\bar\o(\o+\bar\o)+\frac{g^6}{8}\o\bar\o ((\o+\bar\o)^2+2\o\bar\o )\\
     &-&\frac{g^8}{16}\o\bar\o(\o+\bar\o)\bigl((\o+\bar\o)^2+7\o\bar\o \bigr)
     +\frac{g^{10}}{32}\o\bar\o\bigl((\o+\bar\o)^4+16 \o\bar\o ( \o+\bar\o)^2+11(\o\bar\o)^2\bigr)+\dots\nonumber
\eea
implying that $\Lambda(\o, {\bar\o})$ is
\bea
\Lambda &=&\frac{1}{2} 
     -\frac{g^4}{4}(\o+\bar\o)+\frac{g^6}{8} ((\o+\bar\o)^2+2\o\bar\o )\\
     &-&\frac{g^8}{16}(\o+\bar\o)\bigl((\o+\bar\o)^2+7\o\bar\o \bigr)
     +\frac{g^{10}}{32}\bigl((\o+\bar\o)^4+16  ( \o+\bar\o)^2+11(\o\bar\o)^2\bigr)+\dots
\nonumber
\eea

More generally, both the general deformation considered in eq.~(\ref{biAnsatz})
and the function $\Lambda$ have one free coefficient for every fourth power of the field 
strength suggesting that there should exist a one to one map between the two 
functions.
Thus, ${\cal N}=1$ supersymmetry does not seem to rule out any of the solutions with positive 
energy constructed using either \, \sect{BNproc} or more generally \, \sect{genProc}: for every such 
model one may easily find $\Lambda$ (at least perturbatively) and thus construct an action in 
${\cal N}=1$ superspace whose bosonic component reproduces the initial bosonic action. This result 
is not completely surprising; it was shown in~\cite{Kuzenko:2000uh} that all solutions of the bosonic 
NGZ equation have an ${\cal N}=1$ supersymmetric completion. 
Since all relevant solutions of the deformed  self-duality constraint (\ref{defGen}) are solutions of 
the NGZ relation, the same conclusion must apply to them as well.

\subsection{${\cal N} = 2$ supersymmetric non-linear $U(1)$ duality models}

While all actions constructed in earlier sections have an ${\cal N}=1$ supersymmetric extension, 
most of them do not have a known extended supersymmetric counterpart.  It may be also useful to recall here 
  the results of~\cite{BG, RT}, namely that the Born-Infeld action is  unique in that it has  
4 linearly realized and 4 nonlinearly realized supercharges.

The ${\cal N}=2$ global superspace is parametrized by 
$\cZ^A = (x^a, \q^\a_i, {\bar \q}^i_\ad) $, with $i = {1}, {2}$ 
being the $SU(2)$ R-symmetry  index. 
Actions describing the dynamics of ${\cal N}=2$ vector multiplets  are written in terms of the  
(anti) chiral superfield strengths ${\overline \cW}$ and $\cW$ which satisfy the Bianchi identities
\footnote{The derivatives $\cD^{ij}$ and ${\overline \cD}{}^{\,ij}$ are defined as 
$\cD^{ij}=\cD^{i\alpha}\cD^j_\alpha$ and
${\overline \cD}{}^{\,ij}={\overline \cD}{}^{\,i}_{\dot\alpha}{\overline \cD}{}^{j{\dot\alpha}}$.}
\be
\cD^{ij} \, \cW  \, = \,  {\overline \cD}{}^{\,ij} \, {\overline \cW} \ .
\label{n=2bi-i}
\ee
They determine the superfield strength in terms of an unconstrained prepotential $V_{ij}$.   
\be
\cW =  {\overline \cD}^{\, 4} \cD^{ij} \, V_{ij} \, , \qquad \quad 
{\overline \cW} = \cD^{\, 4} {\overline \cD}^{\,ij} \, V_{ij} \ ,
\ee
where ${\overline \cD}^{\, 4}$ is a chiral projector: ${\overline \cD}^{\, i}_\alpha {\overline \cD}^{\, 4} U=0$ for any superfield $U$.

As in the case of ${\cal N}=1$ supersymmetric models one may define, following \cite{Kuzenko:2000uh},  
dual (anti) chiral superfields $\overline {\cal M}$ and ${\cal M}$ as
\be
{\rm i}\,{\cal M} \equiv 4\, \frac{\d }{\d \cW}\,
\cS[\cW , {\overline \cW}]
 \, , \qquad \quad
- {\rm i}\,{\overline {\cal M}} \equiv 4\, 
\frac{\d }{\d {\overline \cW}}\, {\cal S}[\cW , {\overline \cW}]
\label{n=2vd}
\ee
in terms of which the equations of motion are
\be
\cD^{ij} \, {\cal M}  \, = \,  {\overline \cD}^{\,ij} \, \overline {\cal M} \, .
\label{n=2em}
\ee

To construct the ${\cal N}=2$ analog of the NGZ relation we note that, similarly to the 
${\cal N}=1$ setup, the Bianchi identities (\ref{n=2bi-i}) and the equations of motion
(\ref{n=2em}) have the same functional form and are mapped into each other by the infinitesimal 
U(1) duality transformations
\be
\d \cW  \, = \,  \l \, {\cal M} \, , \qquad \quad
\d {\cal M}   \, = \,  -\l \, \cW \, .
\label{n=2dt}
\ee
Considering the fact that the ${\cal N}=2$ NGZ identity should reduce to the 
equation \, (\ref{n=1dualeq}) upon ignoring the fields in the ${\cal N}=1$ chiral 
multiplet, we are left with \cite{Kuzenko:2000uh}
\be
\int {\rm d}^8 \cZ\, 
\Big( \cW^2 + {\cal M}^2 \Big)  \, = \, 
\int {\rm d}^8 {\bar \cZ}\,
\Big( {\overline \cW}^2  +
{\overline {\cal M} }^2 \Big)
\label{n=2dualeq}
\ee
as the only possible ${\cal N}=2$ extension of (\ref{n=1dualeq}). 
Solutions of this equation have not been easy to find. 
The free ${\cal N}=2$ supersymmetric Maxwell action
 \be
{\cal S}_{\rm free} = 
\frac{1}{8}\int {\rm d}^8 \cZ \, \cW^2 +
\frac{1}{8}\int {\rm d}^8{\overline \cZ} \;{\overline \cW}^2 
\label{n=2maxwell}
\ee
satisfies this constraint. The one other known action obeying the constraint (\ref{n=2dualeq}) was discovered by Ketov in~\cite{Ket2}. It is
\be
{\cal S}  \, = \, 
\frac{1}{4}\int {\rm d}^8  \cZ \,  {\cal X} +
\frac{1}{4}\int {\rm d}^8{\overline \cZ} \,{\overline {\cal X}} \, ,
\label{n=2bi}
\ee
where the chiral superfield ${\cal X}$ is a functional 
of $\cW$ and $\overline \cW$ and is a solution of the constraint
\be
{\cal X}  \, = \,  {\cal X} \, {\overline \cD}^{\, 4} {\overline {\cal X}}  \, + \, 
\hf \, \cW^2 \, .
\label{n=2con}
\ee
Upon solving the constraint (\ref{n=2con}), the action becomes~\cite{Ket2,Kuzenko:2000uh, Ketov:2001dq,Bellucci:2001hd} 
\be
S_{N=2}= S_{\rm free} + \int d^4x \, d^8 \theta \; 
{\cal W}^2 \, {\overline{\cal W}}^2\,
{\cal Y}\, (D^4{\cal W}^2, \bar D^4{\overline {\cal W}}^2) + {\cal O}(\partial_\mu {\cal W})
\label{Ketov}
\ee
where ${\cal Y}$ is a Born-Infeld-type functional which in the ${\cal N}=0$ limit reduces to $\Lambda_{\rm BI}(\o,\bar \o)$ in eq.~(\ref{bi-lag}).
%

The system (\ref{n=2bi}), (\ref{n=2con}) was introduced in~\cite{Ket2} as 
the ${\cal N}=2$ generalization of the Born-Infeld action.
In ${\cal N}=1$ language, the ${\cal N}=2$ vector 
multiplet splits into a vector and a chiral ${\cal N}=1$ multiplets. By truncating away the
chiral multiplet the equations above  correctly reproduce the system 
(\ref{gendualaction}), (\ref{GZ4'}) and (\ref{bi-lag}).

The extra terms with derivatives $\partial_\mu {\cal W}$ appear to be required for $N>1$ actions. 
%
%
Moreover, the only solutions presented explicitly in the literature which have manifest $N=2$ supersymmetry and are compatible with the duality condition also have the structure of the BI action but  exhibit additional terms containing space-time derivatives\footnote{This state of affairs appears to be different from the statement \cite{BN} that the extension of the BN construction to a supersymmetric setup does not encounter {\it any} difficulties. It is not clear to us whether this statement refers to minimal or extended supersymmetry. In our discussion there is a fundamental difference between minimal and extended supersymmetry, the former accommodating indeed any solution of the deformed self-duality equation. 
}. 
They also share the property that they are associated with the D3-brane actions
$ 
L_{{\rm D3-brane}} \, = \, 1- \, \sqrt{-\det \big( \eta_{ab} + F_{ab} 
+ \pa_a {\bar \vf} \pa_b \vf \big)} 
$.
It was shown in
\cite{Kuzenko:2000uh} that an ${\cal N}=2$ self-dual action  is given by
\bea
\cS_{\rm BI} &=& \cS_{\rm free}  \, + \,  \cS_{\rm int}
\\
\cS_{\rm int} &=& 
{ 1 \over 8} \,  \int d^{4}xd^8\theta \, 
\cW^2\,{\overline \cW}^2\, \Bigg\{ 1 + 
\hf \, \Big( \cD^4 \cW^2 + {\overline \cD}^{\, 4} {\overline \cW}^2 \Big)\label{5-int}   \\
&+&\frac{1}{4} \, 
\Big( (\cD^4 \cW^2)^2 + ({\overline \cD}^{\, 4} {\overline \cW}^2)^2 \Big)
+ \frac{3}{4}\, (\cD^4 \cW^2)({\overline \cD}^{\, 4} {\overline \cW}^2)
\Bigg\} \non \\
&+&{ 1 \over 24} \,  \int {\rm d}^{12} \cZ \, 
\Bigg\{ {1 \over 3} \cW^3 \Box {\overline \cW}^3 
+\hf (\cW^3 \Box {\overline \cW}^3) {\overline \cD}^{\, 4} {\overline \cW}^2 
+\hf ({\overline \cW}^3 \Box \cW^3) \cD^4 \cW^2 
+{ 1 \over 48} \cW^4 \Box^2 {\overline \cW}^4 \Bigg\} \non \\
&+& \, \cO(\cW^{10}) \, . \non
\eea 
The unique term with no fermionic or space-time derivatives, $\cW^2\,{\overline \cW}^2$,
yields the known $F^4$ term of the Born-Infeld action.  The six-order terms, apart from $\cW^3 \Box {\overline \cW}^3$ terms with space-time derivatives, also correspond to the BI model. This action was confirmed in \cite{Bellucci:2001hd,Ketov:2001dq}.

At the current level of understanding of $N=2$ supersymmetric duality symmetric theories it is not clear yet what role will be played by the BN proposal to deform the twisted self-duality equation. The terms with space-time derivatives  of the superfields are not likely to be generated by the initial deformation of  self-duality equation, unless one allows for 
deformations which contain derivatives of the field strength\footnote{With such a deformation 
it possible that the resulting action is nonlocal (though perturbatively local), as demonstrated in~\cite{BN}
for the case of and $C^2 (dF)^2$ deformation of maximal supergravity.}.

\section{Discussion}
\label{Discussion} 

The question whether duality symmetries of equations of motion survive quantization
and constrain the effective action of the theory is very interesting and with far reaching 
implications both for gravitational and non-gravitational theories. A direct construction based 
on the classical Lagrangian and some number of (perhaps quantum generated) local counterterms
would extend the tally of duality invariant theories and could shed light on the quantum properties 
of the theory. For supergravity theories in general and for ${\cal N}=8$ supergravity in particular 
it may constrain the existence of higher-loop counterterms not immediately amenable to explicit 
calculations. 
In cases in which  only the classical equations of motion are invariant under duality 
transformations (while the action is not), the construction is complicated by the fact that 
simply adding to the action a duality invariant counterterm leads~\cite{Kallosh:2011dp} 
to duality-non-invariant deformed equations of motion and a non-conserved  NGZ duality 
current.

In ref.~\cite{BN}  a procedure, which we have broken into five-steps in~\sect{BNproc}, was suggested such that an action exhibiting a conserved NGZ duality current is constructed if the procedure can be carried out.  This directly follows if the first counterterm/deformation is manifestly duality invariant.  The deformations discussed in ref.~\cite{BN} are assumed to depend on fields transforming linearly under 
duality transformations; in supergravity theories they are the vector fields.  
The action constructed following the BN procedure has infinitely many terms which, in the presence of 
derivatives acting  on the field strengths, may also be nonlocal though local order by order in a weak coupling expansion.

To understand and test this proposal we studied in detail a simple example -- that of nonlinear 
electrodynamics.  We found that, while an action can always be constructed, this action typically 
does not have desirable properties unless one assumes the existence of higher-order deformations 
of a specific form.
In particular, using known results of supersymmetric nonlinear actions for abelian vector multiplets, we 
find that the Bossard-Nicolai action generated by the first ${\cal I}^{(1)}\sim F^4$ deformation of the linear twisted self-duality constraint, may not  have a  supersymmetric generalization beyond 
${\cal N}=1$ supersymmetry.  
To recover the known $N=2$ actions of the BI type, the deformation of the linear twisted self-duality constraint must be modified to include 
%
%
all order terms ${\cal I}^{(n)}\sim F^{4n}$. The generalized construction, extending that of BN, is detailed in section \ref{genProc}. Moreover, for $N>1$ the action must depend of space-time derivatives of the superfields and, correspondingly on space-time derivatives of  $F_{\mu\nu}$. Therefore it is not clear what kind of deformed linear twisted duality constraint will provide the action consistent with $N>1$ supersymmetry and duality.

In the extended supersymmetric case of nonlinear electrodynamics the higher-order counterterms/deformations may be found by simply requiring that the resulting duality-invariant action has more that 8 supercharges.  We believe that a similar requirement will generally restrict the large class of actions allowed by our construction. 

It is possible that, in general, the required higher-order counterterms may be found by simply requiring 
that, order by order in perturbation theory, the action generated by our procedure can be supersymmetrized.  
It is  unclear, however,  whether this requirement is sufficient to generate a correct or unique action.   In an interacting 
theory, the terms found in such a manner may very well be incompatible with those 
generated by standard perturbation theory. It is possible that terms  that are separately invariant 
under supersymmetry transformations may need to be added.

As we have seen, the perturbative deformation of the  linear twisted self-duality constraint suggested in ref.~\cite{BN} requires in addition 
the presence of infinitely many terms to recover the Born-Infeld action. The non-universality 
 ({i.e.} the fact that they 
are not uniquely determined by the first deformation/counterterm and the duality constraint) of 
the higher order terms is somewhat troublesome.  It does not indicate that the BN procedure leads to an unconditional success for all non-linear duality theories. We have also discussed an alternative 
twisted self-duality constraint -- initially suggested by Schr\"odinger  -- which leads to the Born-Infeld action 
while not requiring order by order corrections. The fundamental difference between this approach 
and the perturbative one is that the Schr\"odinger constraint is completely cubic; attempting to reconstruct the 
perturbative deformation of the linear  self-duality constraint necessarily leads to 
terms with non-analytic dependence on $T^-$, as follows from (\ref{cubiccomb1}). The existence of 
two twisted self-duality relations that yield the Born-Infeld action suggests it may be a general feature of 
this construction of duality-invariant actions.

Part of the motivation behind understanding the construction of actions exhibiting non-linear 
duality symmetries is provided by applications to supergravity theories. In maximal four-dimensional 
supergravity it was shown from several standpoints~\cite{Kallosh:1980fi,Howe:1980th}, \cite{Beisert:2010jx}-\cite{Green:2010sp}, 
 that the first  $E_{7(7)}$ duality-invariant potential counterterm may occur at 7 or 8 loops. Supersymmetry considerations
as well as the structure of scattering amplitudes of ${\cal N}=8$ supergravity imply that this counterterm 
necessarily contains terms quartic in vector fields. Assuming that the $E_{7(7)}$ duality symmetry should 
survive quantization,  one is therefore to 
attempt to construct non-linear duality models\footnote{Such models are 
expected to contain arbitrary powers of the vector field strength. Presumably, these terms 
should be related to terms identified in the analysis of~\cite{Beisert:2010jx} as required for 
having vanishing soft-scalar limits for multi-point S-matrix elements.} with maximal supersymmetry and with scalar field dependence which twists nontrivially 
the classical duality constraint. Such models have never been constructed before.
Our generalization of the BN proposal, which accounts for known models of non-linear duality, offers a wide pool of bosonic models among which there may exist one which admits a maximally supersymmetric completion. The nontrivial way in which a supersymmetric Born-Infeld action emerged from such an analysis makes it difficult to conclude, however, that such a model must exist and what is its precise structure and relation to the first counterterm. 
Further detailed analysis is necessary to unravel this issue; along the way to maximal supersymmetry and supergravity we 
may find novel models of nonlinear duality which are interesting in their own right. 

{\it Note added.}  When this paper was finalized we were informed by G.~Bossard and  H.~Nicolai 
 that they have also worked out the Born-Infeld theory in the (Floreanini-Jackiw)-Henneaux-Teitelboim formulation.

\section*{Acknowledgments}
We are grateful to  Z.~Bern, J.~Broedel, G.~Bossard, S.~Ferrara, D.~Freedman, A.~Linde, H.~Nicolai, D.~Sorokin, M.~Tonin and A.~Tseytlin for  stimulating discussions.   We would especially like to thank S.~Ferrara and A.~Tseytlin for insightful comments on an initial draft and S. Ketov and S. Kuzenko for their help in understanding the issues of ${\cal N}=2$ supersymmetry and self-duality.   This work  is supported by the Stanford Institute for Theoretical Physics, NSF grants 0756174, PHY-08-55356, and the A.P.~Sloan Foundation.  

\appendix
\section{Generalization to supergravity}
\label{sugraGenApp}

\subsection{Duality and Supergravity}

The action of the $n$ vector fields of an ${\cal N}>2$ extended supergravity theory is
\be
{\cal L}_{\rm vectors}= i\, \overline {\cal N}_{\Lambda \Sigma} F^{-\Lambda} F^{-\Sigma}+h.c. \ ,
 \ee 
where ${\cal N}_{\Lambda \Sigma}(\phi)$ is a scalar field dependent symmetric matrix. 
The scalar fields $\phi$ parametrize a coset $G/H$ with the theory-specific duality group $G$ and  its subgroup $H$ isomorphic to the R-symmetry group. For ${\cal N}=8$ supergravity $G=E_{7(7)}$ and $H=SU(8)$.
The self-duality constraint derived from (\ref{covConstraint}) is twisted by this matrix and may be written either as a $G$ covariant constraint
\be
 G_\Lambda^+= {\cal N}_{\Lambda \Sigma} F^{+\Lambda}\, , \qquad  G_\Lambda^-= \overline {\cal N}_{\Lambda \Sigma} F^{-\Lambda}\, , 
 \ee
or as an $H$ covariant one
\be
T^+_{AB}=0 \ ,
\label{LinearT}
\ee
where
\be
T^\pm \equiv  h_{\Lambda AB} F^{\pm \Lambda}- f_{AB}^\Lambda G_\Lambda^pm\,
\label{GP}\ee
and where the kinetic term matrix ${\cal N}_{\Lambda \Sigma}(\phi)$ is constructed out of the scalar field-dependent sections of an $Sp(2n_v, \mathbb{R})$ bundle over the $G/H$ coset space $h_{\Lambda AB} $ and $f_{AB}^\Lambda$; they transform in an antisymmetric representation of $H$ -- see~\cite{Gaillard:1981rj, Aschieri:2008ns, Andrianopoli:1996ve} for details.
The equations (\ref{LinearT}) are the supergravity analog of \eqn{maxwellduality}.

An infinitesimal $Sp(2n_v, {\mathbb{R}})$ transformation acts on a duality vector field doublet 
in a real representation exactly as given in \eqn{symplectic}.
Here, as there, $A,  B, C, D$ are the infinitesimal parameters of the transformations, arbitrary real $n_v\times n_v$ matrices satisfying (\ref{symplectic}). The vector kinetic matrix transforms projectively under $Sp(2n_v, \mathbb{R})$
 \be
{\cal N}'= (C+D{\cal N})(A+B{\cal N})^{-1}  \ .
\label{calN} 
\ee

The case of the graviphoton in the absence of scalars and of additional vector fields, $A=D=0$ and $B=-C$,  the $U(1)\sim SO(2)$, follows the Maxwell discussion of \sect{dualityEM} identically.   

In ${\cal N}=8$ supergravity, for $E_{7(7)}$, the NGZ identity requires that the following functional differential equation  be satisfied 
\be
{\delta \over \delta F(y)}\Big (  \delta S- {1\over 4} \int d^4x (\tilde G B G+ \tilde FC F )\Big )=0\,,
\label{const}
\ee
where $\delta S$ is the variation of the action under $E_{7(7)} $  
\be
 \delta S= {\delta S \over \delta F} \delta F + {\delta S \over \delta \phi} \delta \phi \, ,
\label{deltaS}
\ee
and $\delta F$ and $\delta \phi$ are the variations of vectors and scalars, respectively, under $E_{7(7)} $.  
Here the $E_{7(7)}$ symmetry transformations in the real basis for the doublet $(F, G)$ are defined by  an $Sp(2n, \mathbb{R})$ embedding
\begin{eqnarray}\label{E77}
  \left(
                                        \begin{array}{cc}
                                        A&  \; B \\
                                        C & \;  D \\
                                        \end{array}
                                      \right)
=\left(
                                        \begin{array}{cc}
                                         \mbox{Re}\Lambda-\mbox{Re}\Sigma&  \; \mbox{Im} \Lambda+ \mbox{Im} \Sigma \\
                                         - \mbox{Im}\Lambda+ \mbox{Im}\Sigma & \;  \mbox{Re}\Lambda+ \mbox{Re}\Sigma \\
                                        \end{array}
                                      \right)\, 
\end{eqnarray}
 $\Lambda$ are parameters of  $SU(8)$ and $\Sigma$ are the $SU(8)$-orthogonal parameters of $E_{7(7)}$ , which control the familiar infinitesimal shift of scalars $\delta \phi = \Sigma+\cdots$. 
 
\subsection{Modification of procedures}

The modification to the procedures of \sect{BNproc} and \sect{genProc} is actually quite minimal in terms of the algorithms.  What grows in complexity, which may be the reason there are no non-linear examples currently worked out in supergravity, is the complexity of the NGZ identity that must be maintained.  In the ${\cal N}=8$ supergravity case it is actually \eqn{const} which must be satisfied order by order.

\section{Born-Infeld and Bossard-Nicolai Hamiltonians}
\label{HamiltonianAppendix}
In $U(1)$  duality invariant models 
 there is a simple relation between the Lagrangian and the Hamiltonian formulations~\cite{Gibbons:1995cv,Gaillard:1997rt}. The NGZ constraint discussed above can be expressed as a differential equation with  solutions, perturbative in $g^2$, codified in an arbitrary function of one real variable.  
 
 The Lagrangian  can be expressed in terms of  $t=\frac{1}{4} F^2$ and  on $ z=\frac{1}{4} F \tilde{F}$.  We introduce the following (copious) notation to touch the (equally copious) literature
 \bea
 x&=&  \sqrt{t^2+z^2}\, ,\\
 y&=& - {\textstyle \frac{1}{2}}z^2\, ,\\
 Y&=& x^2 \, ,\\
 X&=& t\,.
 \eea
 We can write same Hamiltonian as two different functional forms $H(X,y)=V(X,Y)$.  Similarly we can write the same Lagrangian as two different functional forms $ {\cal L}(t,z)=k(t,x)$.    
 
 The nice relation between $U(1)$ duality-conserving Lagrangians and Hamiltonians is simply
 \be
  {\cal L}(t,z)=k(t,x)=-H(X,y)=-V(X,Y)\,.
 \ee
 These represent general solutions of the differential equation,
 \be
( \partial_t k)^2 - (\partial_x k)^2 = 1
\ee
which is simply another way of writing the NGZ constraint, (c.f.~\eqn{lagrangeNGZ}).

For example, for Maxwell and for Born-Infeld the respective functional forms are simply
\begin{align}
  {\cal L}_{\rm Max}(t, z)&= -t&   {\cal L}_{\rm BI}(t, z)&= -g^{-2} \Big( \sqrt{1+2g^2t- g^4z^2} - 1 \Big)\\
H_{\rm Max}(X, y)&= X&   H_{\rm BI}(X, y)&= g^{-2} \Big( \sqrt{1+2g^2X+2g^4y}-1 \Big)\\
V_{\rm Max}(X, Y)&= X&  V_{\rm BI}(X, Y)&=  g^{-2} \Big( \sqrt{1+2g^2X+g^4 X^2- g^4Y}-1 \Big)
\end{align}

For the BN model (see sections {\ref{BNLagrangianSection}} and {\ref{BNdualitySection}}),  we have

\begin{multline}
\label{BNaction1}
{\cal L}_{\rm BN}(t, z,g^2)= -t + \frac{1}{2} g^2 \left(t^2+z^2\right)  -\frac{1}{2} g^4\, t\, \left(t^2+z^2\right) + \frac{1}{4} g^6 \left(t^2+z^2\right) \left(3 t^2+z^2\right)
-\frac{1}{8} g^8\, t \, \left(t^2+z^2\right) \left(11 t^2+7 z^2\right)  \\
+ \frac{1}{32} g^{10} \left(t^2+z^2\right) \left(  91  t^4+  86 t^2 z^2+11 z^4\right) 
-\frac{1}{8} g^{12} \, t\,  \left(t^2+z^2\right) \left(51 t^4+64 t^2 z^2+17 z^4\right)\\
+\frac{1}{64} g^{14} \left(t^2+z^2\right) \left(969 t^6+1517 t^4 z^2+623 t^2 z^4+43 z^6\right) + \cdots \, .
\end{multline}%
It follows that
\begin{multline}
\label{BNHam}
V_{\rm BN}(X, Y,g^2)= X - \frac{1}{2} g^2 Y  +\frac{1}{2} g^4\, X\, Y - \frac{1}{4} g^6 Y \left(2 X^2+Y\right)
+\frac{1}{8} g^8\, X \, Y \left(4 X^2+7 Y\right) \\  
- \frac{1}{32} g^{10} Y \left(  16X^4  +  64X^2Y  +11 Y^2\right) 
+\frac{1}{8} g^{12} \, X\, Y \left(4 X^4 + 30 X^2 Y + 17 Y^2\right)\\
-\frac{1}{64} g^{14} Y \left(32 X^6 + 400 X^4 Y + 494 X^2 Y^2 + 43 Y^3\right) + \cdots \, .
\end{multline}%

The sign of $g^2$ can be adjusted in the non-covariant procedure through a suitable choice for the first integration constant.  Notice that when we make a choice $g^2=-1$, which is the choice made in ref.~\cite{BN}, we find
\begin{multline}
\label{BNHamgneg}
V_{\rm BN}(X, Y, {\displaystyle g^2=-1})= X + \frac{1}{2}  Y  +\frac{1}{2} \, X\, Y + \frac{1}{4}  Y \left(2 X^2+Y\right)
+\frac{1}{8} \, X \, Y \left(4 X^2+7 Y\right) 
+ \frac{1}{32}  Y \left(  16X^4  +  64X^2Y  +11 Y^2\right) \\
+\frac{1}{8}  \, X\, Y \left(4 X^4 + 30 X^2 Y + 17 Y^2\right)
+\frac{1}{64}  Y \left(32 X^6 + 400 X^4 Y + 494 X^2 Y^2 + 43 Y^3\right) + \cdots\\
= X+\frac{1}{2} \, Y\,( X+ X^2 +X^3+\cdots)+ \frac{1}{4}  Y^2 +\cdots\, . 
\end{multline}%
The last line is in  agreement with ref.~\cite{BN}. It also explains the choice of $g^2=-1$, since it provides a positive definite Hamiltonian at each order. Since the BN solution does not have a closed form expression\footnote{At least not to our knowledge.}, the choice of $g^2=-1$ for the positivity of $H$ means that the quartic deformation of the action has a sign opposite to the BI model.  Note that the BI Hamiltonian is not positive definite at each order, only the closed form expression is positive.



\begin{thebibliography}{10}

\bibitem{Cremmer:1978km}
  E.~Cremmer, B.~Julia, J.~Scherk,
  ``Supergravity Theory in Eleven-Dimensions,''
  Phys.\ Lett.\  {\bf B76}, 409-412 (1978).
  E.~Cremmer and B.~Julia, 
``The SO(8) Supergravity,"
  Nucl.\ Phys.\  B {\bf 159}, 141 (1979).
  B.~de Wit and H.~Nicolai, 
 ``${\cal N}=8$ Supergravity,"
  Nucl.\ Phys.\  B {\bf 208}, 323 (1982).
  B.~de Wit, 
``Properties Of SO(8) Extended Supergravity,"
  Nucl.\ Phys.\  B {\bf 158}, 189 (1979).
  B.~de Wit and D.~Z.~Freedman, 
``On SO(8) Extended Supergravity,"
  Nucl.\ Phys.\  B {\bf 130}, 105 (1977).
  
  

\bibitem{Gaillard:1981rj}
  M.~K.~Gaillard and B.~Zumino, 
``Duality Rotations For Interacting Fields,"
  Nucl.\ Phys.\  B {\bf 193}, 221 (1981).
  
  
   
\bibitem{Kuzenko:2000uh}
  S.~M.~Kuzenko and S.~Theisen, 
  ``Nonlinear selfduality and supersymmetry,"
  Fortsch.\ Phys.\  {\bf 49}, 273 (2001)
  [arXiv:hep-th/0007231].
  S.~M.~Kuzenko and S.~Theisen, 
  ``Supersymmetric duality rotations,"
  JHEP {\bf 0003}, 034 (2000)
  [arXiv:hep-th/0001068].
  
  
\bibitem{Aschieri:2008ns}
  P.~Aschieri, S.~Ferrara and B.~Zumino, 
  ``Duality Rotations in Nonlinear Electrodynamics and in Extended
  Supergravity,"
  Riv.\ Nuovo Cim.\  {\bf 31}, 625 (2008)
  [arXiv:0807.4039 [hep-th]].
  
  
\bibitem{Ferrara:1976iq}
  S.~Ferrara, J.~Scherk, B.~Zumino,
  ``Algebraic Properties of Extended Supergravity Theories,''
  Nucl.\ Phys.\  {\bf B121}, 393 (1977).

\bibitem{Cremmer:1977tt}
  E.~Cremmer, J.~Scherk, S.~Ferrara,
  ``SU(4) Invariant Supergravity Theory,''
  Phys.\ Lett.\  {\bf B74}, 61 (1978).
 

  
  
      \bibitem{BI}
M.~Born and L.~Infeld, ``Foundations of the New Field Theory'', Proc. Roy. Soc. (London) {\bf A144}, 425 (1934).

 \bibitem{Schrodinger}
  E. Schr\"odinger, ``Contributions to Born's New Theory of the Electromagnetic Field'', Proc. Roy. Soc. (London) {\bf A150}, 465 (1935).
  
\bibitem{Cecotti:1986gb}
  S.~Cecotti and S.~Ferrara, 
  ``Supersymmetric Born-Infeld Lagrangians,"
  Phys.\ Lett.\  B {\bf 187}, 335 (1987).
  

  
\bibitem{Gibbons:1995cv}
  G.~W.~Gibbons and D.~A.~Rasheed, 
  ``Electric - magnetic duality rotations in nonlinear electrodynamics,"
  Nucl.\ Phys.\  B {\bf 454}, 185 (1995)
  [arXiv:hep-th/9506035].

 \bibitem{Gaillard:1997rt}
  M.~K.~Gaillard and B.~Zumino, 
``Nonlinear electromagnetic self-duality and Legendre transformations,"
  arXiv:hep-th/9712103.
  M.~Hatsuda, K.~Kamimura and S.~Sekiya, 
  Nucl.\ Phys.\  B {\bf 561}, 341 (1999)
  [arXiv:hep-th/9906103].
  X.~Bekaert, S.~Cucu,
  ``Deformations of duality symmetric theories,''
  Nucl.\ Phys.\  {\bf B610}, 433-460 (2001).
  [hep-th/0104048].
  E.~A.~Ivanov and B.~M.~Zupnik, 
  ``New approach to nonlinear electrodynamics: Dualities as symmetries of
  interaction,"
  Phys.\ Atom.\ Nucl.\  {\bf 67}, 2188 (2004)
  [Yad.\ Fiz.\  {\bf 67}, 2212 (2004)]
  [arXiv:hep-th/0303192].
 

\bibitem{Perry:1996mk}
  M.~Perry, J.~H.~Schwarz,
  ``Interacting chiral gauge fields in six-dimensions and Born-Infeld 
theory,''
  Nucl.\ Phys.\  {\bf B489}, 47-64 (1997).
  [hep-th/9611065].
  

 



\bibitem{Tseytlin:1999dj}
  A.~A.~Tseytlin,
  ``Born-Infeld action, supersymmetry and string theory,''
  In *Shifman, M.A. (ed.): The many faces of the superworld* 417-452.
  [hep-th/9908105]. 
  D.~Brace, B.~Morariu and B.~Zumino, 
  ``Duality invariant Born-Infeld theory," In *Shifman, M.A. (ed.): The many faces of the superworld* 103-110
  arXiv:hep-th/9905218.


  
\bibitem{Ketov:2001dq}
  S.~V.~Ketov, 
  ``Many faces of Born-Infeld theory,"
  arXiv:hep-th/0108189.
  
  
\bibitem{Gates:2001ff}
  S.~J.~J.~Gates and S.~V.~Ketov, 
 ``4-D, ${\cal N}=1$ Born-Infeld supergravity,"
  Class.\ Quant.\ Grav.\  {\bf 18}, 3561 (2001)
  [arXiv:hep-th/0104223].
  S.~M.~Kuzenko and S.~A.~McCarthy, 
  ``Nonlinear selfduality and supergravity,"
  JHEP {\bf 0302}, 038 (2003)
  [arXiv:hep-th/0212039].
  
  
\bibitem{Kallosh:1980fi}
  R.~E.~Kallosh, 
``Counterterms in extended supergravities,"
  Phys.\ Lett.\  B {\bf 99} (1981) 122;
  
\bibitem{Howe:1980th}
  P.~S.~Howe and U.~Lindstrom, 
  ``Higher Order Invariants In Extended Supergravity,"
  Nucl.\ Phys.\  B {\bf 181}, 487 (1981).

\bibitem{Howe:1981xy}
  P.~S.~Howe, K.~S.~Stelle and P.~K.~Townsend,
  ``Superactions,''
  Nucl.\ Phys.\  B {\bf 191}, 445 (1981).



\bibitem{Bern:2007hh}
  Z.~Bern, J.~J.~Carrasco, L.~J.~Dixon, H.~Johansson, D.~A.~Kosower and R.~Roiban, 
  ``Three-Loop Superfiniteness of ${\cal{N}}=8$ Supergravity,"
  Phys.\ Rev.\ Lett.\  {\bf 98}, 161303 (2007)
  [arXiv:hep-th/0702112].
  Z.~Bern, J.~J.~M.~Carrasco, L.~J.~Dixon, H.~Johansson, R.~Roiban,
  ``Manifest Ultraviolet Behavior for the Three-Loop Four-Point Amplitude of N=8 Supergravity,''
  Phys.\ Rev.\  {\bf D78}, 105019 (2008).
  [arXiv:0808.4112 [hep-th]].
  Z.~Bern, J.~J.~M.~Carrasco, H.~Johansson,
  ``Perturbative Quantum Gravity as a Double Copy of Gauge Theory,''
  Phys.\ Rev.\ Lett.\  {\bf 105}, 061602 (2010).
  [arXiv:1004.0476 [hep-th]].
  


  
 
  
\bibitem{Brodel:2009hu}
  J.~Broedel and L.~J.~Dixon, 
 ``$R^4$ counterterm and E7(7) symmetry in maximal supergravity,"
  JHEP {\bf 1005}, 003 (2010)
  [arXiv:0911.5704 [hep-th]];
  H.~Elvang and M.~Kiermaier, 
  ``Stringy KLT relations, global symmetries, and $E_{7(7)}$ violation,"
  arXiv:1007.4813 [hep-th].
  H.~Elvang, D.~Z.~Freedman and M.~Kiermaier,
  ``A simple approach to counterterms in N=8 supergravity,''
  JHEP {\bf 1011}, 016 (2010)
  [arXiv:1003.5018 [hep-th]].


\bibitem{Bossard:2010bd}
  G.~Bossard, P.~S.~Howe and K.~S.~Stelle, 
  ``On duality symmetries of supergravity invariants,"
  JHEP {\bf 1101}, 020 (2011)
  [arXiv:1009.0743 [hep-th]].

\bibitem{Beisert:2010jx}
  N.~Beisert, H.~Elvang, D.~Z.~Freedman {\it et al.}, 
  ``E7(7) constraints on counterterms in ${\cal N}=8$ supergravity,"
  Phys.\ Lett.\   B\, {\bf 694}, 265-271 (2010).
  [arXiv:1009.1643 [hep-th]].


\bibitem{Bossard:2011tq}
  G.~Bossard, P.~S.~Howe, K.~S.~Stelle and P.~Vanhove,
  ``The vanishing volume of D=4 superspace,''
  arXiv:1105.6087 [hep-th].

  
\bibitem{Kallosh:2010kk}
  R.~Kallosh, 
  ``The Ultraviolet Finiteness of ${\cal N}=8$ Supergravity,"
  JHEP {\bf 1012}, 009 (2010)
  [arXiv:1009.1135 [hep-th]].
  R.~Kallosh, 
  ``${\cal N}=8$ Supergravity on the Light Cone,"
  Phys.\ Rev.\  D {\bf 80}, 105022 (2009)
  [arXiv:0903.4630 [hep-th]].
  
   \bibitem{spinorString}
J.~Bjornsson and M.~B.~Green,
``5 loops in 24/5 dimensions,''                                               
JHEP {\bf 1008}, 132 (2010)
[1004.2692 [hep-th]];

\bibitem{Green:2010sp}
  M.~B.~Green, J.~G.~Russo, P.~Vanhove,
  ``String theory dualities and supergravity divergences,''
  JHEP {\bf 1006}, 075 (2010).
  [arXiv:1002.3805 [hep-th]].


\bibitem{BDR}
  Z.~Bern, L.~J.~Dixon and R.~Roiban,
  ``Is N = 8 supergravity ultraviolet finite?,''
  Phys.\ Lett.\  B {\bf 644}, 265 (2007)
  [arXiv:hep-th/0611086].
 
\bibitem{FourLoop}
Z.~Bern, J.~J.~Carrasco, L.~J.~Dixon, H.~Johansson and R.~Roiban,
 ``{\it The Ultraviolet Behavior of N=8 Supergravity at Four Loops},''
Phys.\ Rev.\ Lett.\  {\bf 103}, 081301 (2009)
[0905.2326 [hep-th]].
 
\bibitem{Kallosh:2011dp}
  R.~Kallosh, 
  ``$E_{7(7)}$ Symmetry and Finiteness of ${\cal N}=8$ Supergravity,"
  arXiv:1103.4115 [hep-th].
  R.~Kallosh, 
  ``${\cal N}=8$ Counterterms and $E_{7(7)}$ Current Conservation,"
  JHEP {\bf 1106}, 073 (2011)
  [arXiv:1104.5480 [hep-th]].
 
\bibitem{BN}
  G.~Bossard and H.~Nicolai, 
 ``Counterterms vs. Dualities,"
  arXiv:1105.1273 [hep-th].
  
\bibitem{Bossard:2010dq}
  G.~Bossard, C.~Hillmann and H.~Nicolai, 
  ``E7(7) symmetry in perturbatively quantised ${\cal N}=8$ supergravity,"
  JHEP {\bf 1012}, 052 (2010)
  [arXiv:1007.5472 [hep-th]].
  C.~Hillmann, 
  ``E7(7) invariant Lagrangian of d=4 ${\cal N}=8$ supergravity,"
  JHEP {\bf 1004}, 010 (2010)
  [arXiv:0911.5225 [hep-th]].
  
\bibitem{Pasti:1996vs}
  P.~Pasti, D.~P.~Sorokin and M.~Tonin,
``On Lorentz invariant actions for chiral p forms,''
  Phys.\ Rev.\  D {\bf 55}, 6292 (1997)
  [arXiv:hep-th/9611100].
  
\bibitem{DePol:2000re}
  G.~De Pol, H.~Singh and M.~Tonin,
  ``Action with manifest duality for maximally supersymmetric six-dimensional
supergravity,''
  Int.\ J.\ Mod.\ Phys.\  A {\bf 15}, 4447 (2000)
  [arXiv:hep-th/0003106].


\bibitem{Bunster:2011aw}
  C.~Bunster and M.~Henneaux, 
 ``$Sp(2n, \mathbb{R})$ electric-magnetic duality as {\em off-shell} symmetry
of interacting electromagnetic and scalar fields,"
  arXiv:1101.6064 [hep-th].
  
\bibitem{Tseytlin:1990nb}
  A.~A.~Tseytlin,
  ``Duality Symmetric Formulation Of String World Sheet Dynamics,''
  Phys.\ Lett.\  {\bf B242}, 163-174 (1990).
  
\bibitem{Tseytlin:1990va}
  A.~A.~Tseytlin,
  ``Duality symmetric closed string theory and interacting chiral scalars,''
  Nucl.\ Phys.\  {\bf B350}, 395-440 (1991).

\bibitem{Floreanini:1987as}
  R.~Floreanini, R.~Jackiw,
  ``Selfdual Fields as Charge Density Solitons,''
  Phys.\ Rev.\ Lett.\  {\bf 59}, 1873 (1987).
  
  
\bibitem{Henneaux:1988gg}
  M.~Henneaux, C.~Teitelboim,
  ``Dynamics Of Chiral (selfdual) P Forms,''
  Phys.\ Lett.\  {\bf B206}, 650 (1988).


\bibitem{Schwarz:1993vs}
  J.~H.~Schwarz, A.~Sen,
  ``Duality symmetric actions,''
  Nucl.\ Phys.\  {\bf B411}, 35 (1994).
  [hep-th/9304154].  

\bibitem{Andrianopoli:1996ve}
  L.~Andrianopoli, R.~D'Auria and S.~Ferrara, 
  ``U duality and central charges in various dimensions revisited,"
  Int.\ J.\ Mod.\ Phys.\  A {\bf 13}, 431 (1998)
  [arXiv:hep-th/9612105].
  
  
  
\bibitem{Chemissany:2006qd}
  W.~A.~Chemissany, J.~de Jong and M.~de Roo, 
  ``Selfduality of non-linear electrodynamics with derivative corrections,"
  JHEP {\bf 0611}, 086 (2006)
  [arXiv:hep-th/0610060].

  
\bibitem{ArkaniHamed:2008gz}
  N.~Arkani-Hamed, F.~Cachazo, J.~Kaplan,
  ``What is the Simplest Quantum Field Theory?,''
  JHEP {\bf 1009}, 016 (2010).
  [arXiv:0808.1446 [hep-th]].
    
\bibitem{BG} 
J.~Bagger and A.~Galperin, 
``A new Goldstone multiplet for partially broken supersymmetry,"
Phys.\ Rev.\ {\bf D55} (1997) 1091
[hep-th/9608177].



\bibitem{RT}
M.~Ro\v{c}ek and A.A.~Tseytlin, 
``Partial breaking of global D = 4 supersymmetry, 
constrained  superfields, and 3-brane actions,"
Phys.\ Rev.\ {\bf D59}, 106001 (1999)
[hep-th/9811232].


\bibitem{Ket2}
S.V.~Ketov, 
``Born-Infeld-Goldstone superfield 
actions for gauge-fixed D-5 and D-3  branes in 6d,"
Nucl.\ Phys.\  {\bf B553}, 250 (1999) 
[hep-th/9812051].

\bibitem{Bellucci:2001hd}
  S.~Bellucci, E.~Ivanov, S.~Krivonos,
  ``Towards the complete N=2 superfield Born-Infeld action with partially broken N=4 supersymmetry,''
  Phys.\ Rev.\  {\bf D64}, 025014 (2001).
  [hep-th/0101195].
  E.~Ivanov,
  ``Towards higher N superextensions of Born-Infeld theory,''
  Russ.\ Phys.\ J.\  {\bf 45}, 695-708 (2002).
  [hep-th/0202201].

\end{thebibliography}
\end{document}